\documentclass{aa}  

\usepackage{array}
\usepackage{graphicx}
\usepackage{txfonts}
\usepackage{multirow}
\usepackage{amsmath}
\usepackage{txfonts}
\usepackage{xcolor}
\usepackage{graphicx}
\usepackage{bm}
\usepackage{float}

\usepackage{natbib}
\usepackage{booktabs}
\usepackage{fixltx2e}

\usepackage{booktabs}

\usepackage{url}
\usepackage{hyperref}
\hypersetup{
   colorlinks=true,       
   linkcolor=blue,       
   citecolor=blue,        
   urlcolor=blue         
}

\usepackage[normalem]{ulem}

\usepackage{fontawesome5}

\usepackage{pgfplots} 
\usetikzlibrary{external}
\makeatletter
\newcommand{\github}[1]{
   \href{#1}{\faGithubSquare}
}
\makeatother

\newcommand{\HI}{\ifmmode \mathrm{\ion{H}{I}} \else \ion{H}{I} \fi}

\newcommand{\IHI}{\ifmmode I_{{\mathrm{H}} \, \mathrm{I}} \else $I_{{\mathrm{H}} \, \mathrm{I}}$\fi} 

\def\GHz{\ifmmode $\,GHz$\else \,GHz\fi}
\def\MJysr{\ifmmode \,$MJy\,sr\mo$\else \,MJy\,sr\mo\fi}
\def\microns{\ifmmode \,\mu$m$\else \,$\mu$m\fi}

\def\kms{\ifmmode $\,km\,s$^{-1}\else \,km\,s$^{-1}$\fi}

\begin{document} 

\title{From few to many maps: a fast map-level emulator for extreme augmentation of CMB systematics datasets}
   
   \author{P. Campeti \inst{1,2},
   J.-M. Delouis \inst{3},
   L. Pagano \inst{1,4,5},
   E. Allys \inst{6},
   M. Lattanzi \inst{1}, and 
   M. Gerbino \inst{1,4}
   }
    
   \institute{
   \inst{1} INFN Sezione di Ferrara, Via Saragat 1, 44122 Ferrara, Italy \\
   \inst{2} ICSC, Centro Nazionale “High Performance Computing, Big Data and Quantum Computing” \\
   \inst{3} Laboratoire d’Océanographie Physique et Spatiale (LOPS), Univ. Brest, CNRS, Ifremer, IRD, 29200 Brest, France \\
   \inst{4} Dipartimento di Fisica e Scienze della Terra, Università degli Studi di Ferrara, via Saragat 1, I-44122 Ferrara, Italy \\
   \inst{5} Institut d’Astrophysique Spatiale, CNRS, Univ. Paris-Sud, Université Paris-Saclay, Bât. 121, 91405 Orsay cedex, France \\
   \inst{6} Laboratoire de Physique de l’Ecole Normale Supérieure, ENS, Univ. PSL, CNRS, Sorbonne Univ.,
Univ. Paris Cité, 75005 Paris, France
   }

   \date{}
 
  \abstract 
   {Generating massive sets of end-to-end simulations of time-ordered data (TOD) for Monte Carlo analyses in Cosmic Microwave Background (CMB) experiments typically incurs in exceedingly high computational costs.}
   {To address this challenge, we introduce a novel, fast, and efficient generative model built upon scattering covariances, the most recent iteration of the scattering transforms statistics. This model is designed to augment by several orders of magnitude the number of map simulations in datasets of computationally expensive Cosmic Microwave Background (CMB) instrumental systematics simulations, including their non-Gaussian and inhomogeneous features. Unlike conventional neural network-based algorithms, this generative model requires only a minimal number of training samples, making it highly compatible with the computational constraints of typical CMB simulation campaigns. While our primary focus is on spherical data, the framework is inherently versatile and readily applicable to 1D and 2D planar data, leveraging the localized nature of scattering statistics.}
   {We validate the method using realistic simulations of CMB systematics, which are particularly challenging to emulate, and perform extensive statistical tests to confirm its ability to produce new statistically independent approximate realizations.}
  {Remarkably, even when trained on as few as 10 simulations, the emulator closely reproduces key summary statistics—including the angular power spectrum, scattering coefficients, and Minkowski functionals—and provides pixel covariance estimates with substantially reduced sample noise compared to those obtained without augmentation.}
   {The proposed approach has the potential to shift the paradigm in simulation campaign design. Instead of producing large numbers of low- or medium-accuracy simulations, future pipelines can focus on generating a few high-accuracy simulations that are then efficiently augmented using such generative model. This promises significant benefits not only for current and forthcoming cosmological surveys such as \textit{Planck}, \textit{LiteBIRD}, Simons Observatory, CMB-S4, Euclid and Rubin-LSST, but also for diverse fields including oceanography and climate science. We make both the general framework for scattering transform statistics \href{https://github.com/jmdelouis/HealpixML}{\texttt{HealpixML} \faGithub} and the emulator \href{https://github.com/pcampeti/CMBSCAT/}{\texttt{CMBSCAT} \faGithub} available to the community.}


   \authorrunning{Campeti et al.}

   \titlerunning{Map-level emulator for extreme augmentation of small CMB systematics datasets}

   \maketitle

\section{Introduction}\label{sec:intro}

Cosmic Microwave Background (CMB) experiments face the significant challenge of generating massive sets of end-to-end simulations of time-ordered data (TOD) for Monte Carlo analyses. These simulations incur exceedingly high computational costs, often surpassing \(\mathcal{O}(100)\) million CPU hours \citep{Planck:2015txa}. Despite their complexity and expense, these datasets are essential for accurately accounting for all relevant instrumental systematic effects in the statistical inference of cosmological parameters. This process is crucial at both the experiment-design and data-exploitation stages, as exemplified by missions such as the \textit{Planck} satellite \citep[launched in 2009,][]{Planck:2018vyg} and the upcoming \textit{LiteBIRD} satellite \citep[scheduled for launch in the early 2030s,][]{LiteBIRD:2022cnt}. 

The computational demands vary according to the specific experiment, as well as factors such as the scanning strategy, the number of detectors simulated, the chosen simulation framework, and the systematic effects included. Beyond CMB experiments, similar challenges arise also in other forthcoming cosmological surveys, such as Rubin-LSST and Euclid.  In non-cosmological fields, particularly in climate and ocean sciences, simulating spherical datasets can be even more computationally intensive  -- see e.g. \citet{CMPI6_Cost} for the climate Coupled Model Intercomparison Project in its sixth phase (CMIP6) effort -- with some applications requiring up to \(\mathcal{O}(10^3)\) million CPU hours.

Due to their high computational costs, only a limited number of end-to-end simulations—typically a few hundreds—are available for cosmological inference. For example, the most recent \textit{Planck} data release included just 400 simulated maps \citep{Planck:2020olo, Tristram:2021tvh} with simplified instrumental errors to make it computable. However, recent studies \citep{Beck:2022efr} suggest that this number may be insufficient for future CMB experiments. Achieving high-accuracy estimation of cosmological parameters, such as the tensor-to-scalar ratio, may for instance require at least \(\mathcal{O}(10^{4-5})\) simulations. This would correspond to an overwhelming computational cost of approximately \(\mathcal{O}(10^{4-6})\) million CPU hours—far exceeding the typical resource allocations for such scientific projects.
Furthermore, machine learning applications to CMB data, such as direct cosmological parameter extraction from CMB maps using simulation-based inference \citep{Wolz:2023gql}, would demand even larger simulation sets (\(\sim \mathcal{O}(10^{5-6})\) samples) to construct suitable training datasets. 

To address these challenges, we propose a novel approach to dramatically reduce the computational cost of simulation campaigns by augmenting a dataset of few end-to-end map simulations. We show that this data augmentation can be achieved with maximum entropy generative models built from the scattering transform statistics (ST).  
Such models allow to generate approximate realization of a process conditioned on an intermediate low-dimensional (i.e. latent) representation of the target field statistics \citep{Bruna2019, Allys:2020vld}. Specifically, here we adopt as latent representation the latest iteration of ST, the scattering covariance \citep[hereafter SC ][]{Cheng:2023imk}. In contrast to other generative models purely based on neural networks -- such as Generative Adversarial Networks (GANs) \citep{goodfellow_2014}, Variational Auto-Encoders (VAE) \citep{variational_autoencoders_2013} or diffusion models \citep{diffusion}-- with SC we can generate new samples from very small ``training'' datasets, or even from a single realization \citep[i.e. ``one-shot'' or ``few-shots learning'', see e.g.][]{Jeffrey:2021fcg, Mousset}. The concepts and tools related to these generative models also allowed the development of efficient component separation algorithms, which can be applied on a limited number of data, and even directly on observations \citep{Delouis:2022yyt, Auclair2023aej}.

SC \citep[also known as scattering spectra, ][]{Cheng:2023imk} represents the last iteration of the ST statistics, a family of very efficient and powerful summary statistics originally proposed in \citet{mallat_group_2012, bruna2013invariant} to extract information from highly non-Gaussian signals.
The general features of ST statistics are the use of convolution with fixed wavelet filters to separate a field into its individual scales, before evaluating the interaction between these scales using non-linear operations, possibly in an iterative fashion. These statistics include the Wavelet Scattering Transforms \citep[WST, ][]{bruna2013invariant, allys_rwst_2019}, the Wavelet Phase Harmonics \citep[WPH, ][]{zhang_mallat_19, Allys:2020vld}, and the Scattering Covariances/Scattering Spectra \citep{Morel:2022qtr, Cheng:2023imk}. Crucially, ST requires no training because its wavelet kernels are fixed, enabling augmentation of a small training set of tens or hundreds of end-to-end simulations. Moreover, because the wavelet kernels are not learned, the model is more interpretable and has generally fewer free parameters than, for instance, a convolutional neural network (CNN).

ST has proven effective across a wide range of applications \citep[see e.g. ][for a review]{Cheng:2021xdw} and recently made a breakthrough in astrophysics and cosmology, excelling in tasks such as parameter estimation \citep{Cheng:2020qbx, Eickenberg:2022qvy, Valogiannis:2021chp, DES:2023qwe, Valogiannis:2023mxf, SimBIG:2023gke}, classification \citep{bruna2013invariant}, statistical description of complex structures \citep{allys_rwst_2019, Regaldo-SaintBlancard:2020dlb} and component separation \citep{Delouis:2022yyt, regaldo-saint_blancard_new_2021, Auclair2023aej}. 

Moreover, ST has provided a powerful latent representation, enabling the development of maximum entropy generative models that can efficiently sample new approximate realizations (synthetic data or ``emulations'') of homogeneous\footnote{Throughout this work we call a field homogeneous when its statistical properties do not depend on position, i.e. they are assumed to be translation invariant; when this symmetry is broken we refer to the field as inhomogeneous.\label{footnote:footnote1}} astrophysical and cosmological fields, even from a limited number of simulations \citep{Allys:2020vld, Jeffrey:2021fcg, Regaldo23, Price:2023dpj, Mousset}. 
The generation of new synthetic data is typically achieved through gradient descent methods, exploiting statistical mechanics ideas to perform sampling, such as the microcanonical gradient descent \citep{Bruna2019, mean_field}.  Combining ST with gradient descent sampling offers a highly promising approach to address the issues of computational efficiency in statistical inference presented above. It enables production of approximate new samples and therefore augmentation of small ensembles of end-to-end simulations at a fraction of the cost of a full simulation campaign. The ST framework has been successfully applied also to spherical datasets, showcasing its effectiveness in tasks such as component separation \citep{Delouis:2022yyt} and generative modeling \citep{Mousset}, with the latter being the main focus of this paper.

In this work, we present a case study using a full-sky simulation dataset that realistically models the instrumental systematics encountered in CMB satellite experiments to demonstrate the augmentation procedure. We test how the size of the ``training'' dataset impacts the emulations produced by the augmentation process, assuming that only a very limited number --ranging from a few tens to a few hundred-- end-to-end simulations are available.
We validate the generative model (hereafter referred to as the emulator) through an extensive set of visual and statistical tests, requiring that the emulated dataset captures both the true data features (i.e. is not biased compared to the true one) and their variability (i.e. has the same variance).

A fast and efficient map-level emulator offers a number of practical advantages in many aspects of data analysis in cosmology.
Firstly, generating a sufficient number of simulations for cosmological inference can be resource-intensive task, and often one must deal with a limited number of simulations, which can introduce significant sample variance, particularly in empirical covariance matrices. While various approaches have been proposed to mitigate this issue \citep[e.g.,][]{Sellentin:2015waz}, recent studies indicate that certain effects remain difficult to capture without substantially more simulations \citep{Beck:2022efr}.  
In this context, we explore how the augmentation framework can mitigate sample noise in pixel-pixel covariance matrices when the simulation budget is limited to only a few tens or hundreds of expensive end-to-end simulations. By leveraging this framework, current and future CMB surveys such as \textit{Planck} or \textit{LiteBIRD} could better exploit existing data and reduce the non-optimal error bars that often arise from using too few simulations. Moreover, having the ability to generate realistic surrogates from small datasets could shift the current paradigm in many simulation campaigns, allowing to focus both human and computational effort in producing few highly realistic simulations, instead of many simplified ones.  
Beyond the obvious computational budget and environmental considerations, reducing simulation cost is also crucial for enabling faster prototyping of many design choices (e.g., in experiment optimization, foreground modeling, and component separation). 
A fast emulator enables important applications also in simulation-based inference \citep{Cranmer} for cosmological parameters, which is crucial in case the likelihood is unknown or too expensive to evaluate and/or complex and hard to characterize systematics and foregrounds residuals are present \citep{Wolz:2023gql}. 

While the main focus in this paper is applying the model to data on the sphere in \texttt{HEALPix} pixelization \citep{healpix2005}, we emphasize that our framework is inherently versatile and readily applicable to 2D planar and 1D data, leveraging the localized nature of the wavelet transform used to compute the ST statistics. Therefore, our framework can be directly applied to simulated or observed data from current and future CMB ground-based experiments, for instance to flat-sky tiles analysed by BICEP/Keck \citep{BICEP:2021xfz} or to partial-sky \texttt{HEALPix}-pixelized maps from the Small-Aperture-Telescopes (SATs) of Simons Observatory (SO) \citep{SimonsObservatory:2018koc} and CMB-S4 \citep{CMB-S4:2016ple}. By contrast, maps delivered in equirectangular (CAR) grids—such as those produced for SO Large-Aperture-Telescope (LAT) survey—require extra pre-processing steps because the latitude-dependent pixel area of CAR grids can mimic genuine statistical inhomogeneity if left uncorrected.

The paper is organized as follows. In Section~\ref{sec:simulations} we describe how we build the initial ensemble end-to-end of simulations. Section~\ref{sec:algorithm} offers a concise overview of the SC statistics, the generative model, and the data augmentation algorithm, along with a description of the software implementations. Section~\ref{sec:benchmark} provides an estimate of the computational cost of the emulator in various contexts and compares it to a typical full-sky simulation campaign (e.g., for \textit{Planck} or \textit{LiteBIRD}). In Section~\ref{sec:results}, we evaluate the performance of the emulator on two case studies, one trained on just 10 simulations and another on 100 ones. Finally, in Section~\ref{sec:conclusions}, we summarize our findings and discuss future research directions.

\section{Simulations}\label{sec:simulations}

\begin{figure*}[!ht]
\centering
\includegraphics[width=1.\textwidth]{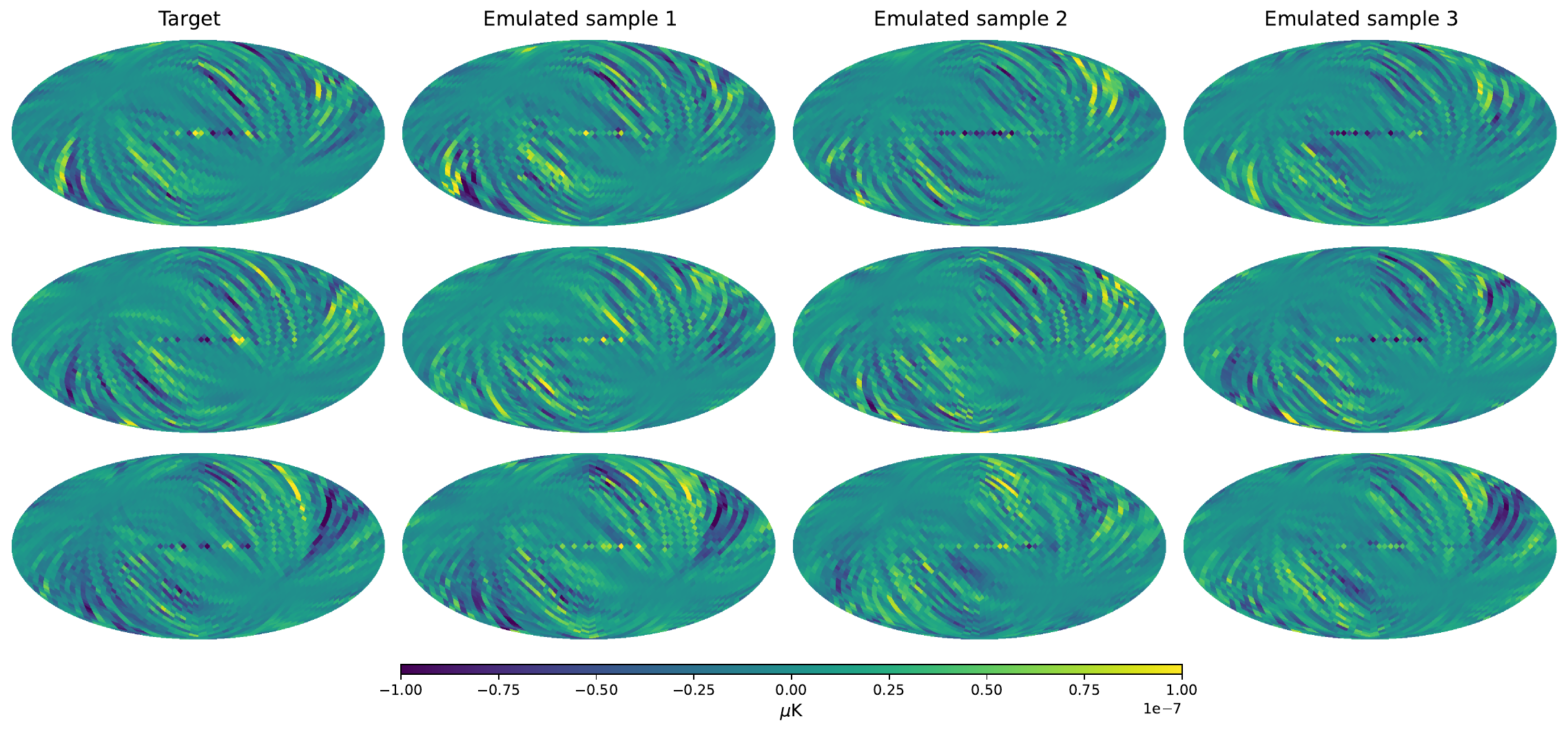}
\caption{Side-by-side comparison of an input $Q$ map (from the moderate size input dataset, labeled ``Target'', in Galactic coordinates) and three corresponding emulated samples synthesized from that input. The leftmost column displays the input map, the other columns present three distinct emulated samples obtained from our generative model, with each row showing a different input target map. The $U$ map (not shown here) shows very similar behaviour. See Sections \ref{sec:simulations} and \ref{sec:sub_results} for further details. 
}
\label{fig:maps}
\end{figure*}

We demonstrate our methods on a very challenging scenario for SC emulation: highly inhomogeneous\footnote{See footnote \ref{footnote:footnote1} for definition of ``inhomogeneous'' in this context.} maps arising from a variable gain systematic, a common instrumental artifact in satellite CMB experiments (e.g., \textit{Planck}). We consider in the following maps containing only such systematic effect, no white or other kind of noises, no CMB signal or lensing and no foregrounds.

We generate 10,000 end-to-end temperature and polarization \(Q\)/\(U\) simulations using the publicly available \texttt{litebird\_sim}\footnote{\href{https://github.com/litebird/litebird_sim}{https://github.com/litebird/litebird\_sim}} time-ordered data (TOD) simulation and map-making toolkit. 
Specifically, we simulate four detectors, arranged in two orthogonally polarized pairs, with a relative orientation of 45 degrees between the pairs. The detectors follow a \textit{Planck}-like scanning strategy \citep{Tauber:2010opu}, and we introduce random gain miscalibration by adding random Gaussian fluctuations at the TOD level.
Each fluctuation is applied to chunks of 100,000 samples ($\sim 4$ hours) with a standard deviation of 0.0001. The resulting maps are produced at \texttt{HEALPix} resolution \(N_{\mathrm{side}}=128\), then downgraded to \(N_{\mathrm{side}}=16\) to decrease computational resources consumption, since we do not expect our conclusions to vary significantly with resolution. The leftmost column of Figure~\ref{fig:maps} (labeled “Target”) shows three example realizations from these simulations.

The variable gain systematic, coupled with the scanning strategy, induces large-scale ``stripes'' features, mainly dominated by dipole leakage \citep[similar to those shown in ][]{Delouis:2019bub}. Because the gain drift we inject depends on the scanning pattern, the resulting maps are strongly inhomogeneous, i.e. their pixel-to-pixel statistics vary across the sky, unlike a homogeneous CMB realization whose statistics depend only on angular separation.
In this work, we validate the emulator on maps containing such highly inhomogeneous structures, which are challenging to emulate with the SC statistics. We discuss in more detail these challenges in Section~\ref{sec:algorithm}. We stress, however, that the method is not limited to this specific systematic. Any effect that imprints spatially varying or non-Gaussian signatures at the map level (e.g. correlated $1/f$ noise, residual foregrounds after component separation, beam asymmetries, bandpass mismatch, etc.) manifests itself as structure in the SC space and could therefore be captured by the same procedure, even if the most complex processes may require some fine-tuning.
The performance of the emulator is expected to improve (or at least not degrade) for systematics that are closer to homogeneous or Gaussian, because the optimization landscape becomes simpler.

\section{Emulator}
\label{sec:algorithm}

\subsection{Basics of generative modelling}\label{basics_of_generative_modelling}
A generative model of a random vector $X$ is an operator $\widehat{G}$ that transforms a Gaussian white noise random vector $Z$ into a model $\widehat{X}=\widehat{G}(Z)$ of $X$ \citep{Ng_jordan}.
In other words, such a model is capable of producing new approximate synthetic observables that realistically approximate the true ones and preserve their underlying probability distribution.

In a broad sense, also simulation can be viewed as a specific type of generative model in which the time evolution (i.e. dynamics) of a physical system is explicitly encoded, controlled by a prescribed set of parameters \citep[see e.g.][]{Price:2023dpj}. Given a set of initial conditions, simulations evolve the system forward in time and return a synthetic observable at a desired point in this evolution. For cosmological observations such as CMB ones, a simulator (or simulation framework) takes an initial configuration of the observing instrument and produces a time series (TOD), i.e. a synthetic sky observation spanning a chosen duration.

By contrast, emulation is a generative model that creates an approximate mapping between the initial conditions and the resulting synthetic observables. In an end-to-end emulation approach, this mapping is learned directly and typically requires large training datasets, the size of which depends on the complexity and dimensionality of the system. When datasets are insufficient for direct end-to-end emulation, one can represent the data using a  lower-dimensional summary statistic—often referred to as a latent representation—and then let the generative model learn the mapping from this reduced representation. Although some information is inevitably lost when using such latent representation, the benefit is a greatly reduced need for extensive training data. The surjectivity of the latent representation introduces variability in the emulations, since several observables can in principle correspond to a single latent vector. 

In the following, we identify two critical requirements for a successful emulator: predictiveness and representativeness. Predictiveness ensures the emulator can reproduce true data features, for instance Galactic foregrounds and localized noise or systematics structures in the maps, or specific patterns in power spectra. Representativeness ensures instead that the emulator can produce (at least approximately) statistically independent realizations that are representative of the true underlying data distribution. Predictiveness and representativeness are tied to the concepts of bias and variance in statistical modeling, where good predictiveness is equivalent to reducing bias, while good representativeness links to capturing variance.

 In cosmology, machine learning algorithms such as GANs \citep{goodfellow_2014} and VAEs \citep{variational_autoencoders_2013} have been successfully employed for end-to-end emulation of relevant observables \citep[][]{Rodriguez:2018mjb,Chardin:2019euc}. However, these networks require large, expensive simulation datasets for initial training. Since the present work deals with relatively small to moderate training sets of CMB data—represented as a pixelized map (via the \texttt{HEALPix} scheme)—we instead leverage the SC statistics.

\subsection{Scattering covariance (SC)}\label{sec:scattering_covariance}

Scattering covariance (SC) is the latest iteration of the scattering transform (ST), originally developed by \citet{mallat_group_2012} and \citet{bruna2013invariant} to provide a more interpretable framework for understanding CNNs \citep[see e.g.][for a review]{Cheng:2021xdw}. Initially applied to one-dimensional signals \citep{Morel:2022qtr}, SC has since been extended to two-dimensional planar fields \citep{Cheng:2023imk} and more recently generalized to data on the sphere \citep{Mousset}.  SC operates by separating an input field \(I\) into different scales by convolving it with a family of complex-valued wavelet filters \(\Psi_\lambda\), where \(\lambda = (j, \gamma)\) specifies both the scale \(j\) and orientation \(\gamma\). After each convolution, SCs are  computed by applying modulus to characterize the interaction between scales and, for higher-order coefficients, computing also covariances among moduli of wavelet-convolved fields to capture additional cross-scale couplings. In this work, we focus on four specific coefficients, as used in \citet{Cheng:2023imk, Mousset}:
\begin{equation}\label{eq:scat_coeff}
\begin{aligned}
 & S_1^{\lambda_1} = \Bigl\langle \bigl| I * \Psi_{\lambda_1} \bigr|\Bigr\rangle\,,  \\
 & S_2^{\lambda_1} = \Bigl\langle \bigl| I * \Psi_{\lambda_1} \bigr|^2\Bigr\rangle\,,  \\
 & S_3^{\lambda_1,\lambda_2} = \mathrm{Cov}\,\Bigl[\,I * \Psi_{\lambda_1},\, \bigl| I * \Psi_{\lambda_2}\bigr| * \Psi_{\lambda_1}\Bigr]\,, \\
 & S_4^{\lambda_1,\lambda_2,\lambda_3} = \mathrm{Cov}\,\Bigl[\bigl| I * \Psi_{\lambda_3}\bigr| * \Psi_{\lambda_1}\,, \bigl| I * \Psi_{\lambda_2}\bigr| * \Psi_{\lambda_1}\Bigr]\,,
\end{aligned}
\end{equation}
where \(*\) denotes convolution, covariances are defined as $\text{Cov}[X,Y] = \langle XY^* \rangle - \langle X \rangle \langle Y^* \rangle
$ for two complex fields $X$ and $Y$ and \(\langle\,\cdot\rangle\) indicates a spatial average.
The first two coefficients measure the first- and second-order moments of single-scale responses, while the latter two encode two- and three-scale couplings, respectively. We use in the following also cross SC statistics computed on two different input fields: more details are reported in Appendix~\ref{sec:cross_SC}.

In all results shown in this paper, we compute wavelet transforms by convolution with a family of dyadic\footnote{The family of wavelets is built by dilating and rotating a mother wavelet. If the dilation factor in the scales is 2 the wavelet family is named dyadic.} wavelets in pixel space on the \texttt{HEALPix} sphere. The specific wavelet filters used in this work, together with details on our specific implementation, are reported in Appendix \ref{sec:oriented_wavel}. The range of scales probed by the wavelets is given by $J_{\rm min} \le j \le J_{\rm max}$ where $J_{\rm min}\ge 0$ and $J_{\rm max}=\log{N_{\rm side}}/\log{2}$, while the number of scales is given by $J = J_{\rm max} - J_{\rm min} +1$. The angular resolution is given in terms of the  number $R$ of orientations. In this work, we adopt for the wavelet filters complex kernels with size $3\times 3$ pixels, probing $J=4$ different scales and $R=4$ orientations. 

As for the normalization of the SC coefficients in Eq.~\ref{eq:scat_coeff} -- which is necessary to avoid issues while optimizing coefficients varying over several orders of magnitude -- we adopt a different convention compared to \citet{Cheng:2023imk} and \citet{Mousset}, and normalize by the $S_2$ coefficient of the same field:
\begin{equation}
    \bar{S}_1^{\lambda_1} = \frac{S_1^{\lambda_1}}{\sqrt{S_2^{\lambda_1}}}, \quad 
    \bar{S}_3^{\lambda_1,\lambda_2} = \frac{S_3^{\lambda_1,\lambda_2}}{\sqrt{S_2^{\lambda_2} S_2^{\lambda_2}}}, \quad 
    \bar{S}_4^{\lambda_1,\lambda_2,\lambda_3} = \frac{S_4^{\lambda_1,\lambda_2,\lambda_3}}{\sqrt{S_2^{\lambda_2} S_2^{\lambda_3}}},\notag
\end{equation}
while $S_2$ is left unnormalized.
The total number of parameters in the SC coefficients amounts to 1472, subdivided in 16 parameters for $S_1$ and $S_2$, 160 parameters for $S_3$ and 1280 parameters for $S_4$.

In principle, the SC model is designed for homogeneous fields, allowing one to estimate higher-order statistics even from a single observation  \citep[e.g.,][]{Jeffrey:2021fcg, Cheng:2023imk}. In our inhomogeneous scenario, additional information to catch the localized structures at large scales is required, which we find particularly efficient to supply by introducing orientation-informed wavelets: we define the SC statistics relative to some primary local orientation (in our case relative to the scanning strategy stripes), which is the orientation with the largest amplitude at each pixel and scale in the wavelet-convolved maps. This allows to precompute a $4\times 4$ (since in our case we deal with $R=4$ wavelet orientations) rotation matrix for each pixel, which rotates the phase of wavelet-convolved maps (before taking the modulus) into the primary orientation basis. The SC statistics computed after this rotation will have amplitudes reflecting the alignment with the primary orientation. We describe in more detail the orientation-informed wavelets in Appendix~\ref{sec:oriented_wavel}.  

We also find that the sharp oscillations -- especially at smaller scales -- in the angular power spectra featured by the specific systematics we consider in this paper (see Fig.~\ref{fig:validation}), are hard to reproduce correctly based only on the SC statistics. In fact, while the first two scattering coefficients can be interpreted as binned power spectra, the features we would like to resolve in the power spectra in this specific case are below the binning resolution introduced by the wavelet convolution. While in principle it is possible to solve this issue by changing the scaling of the wavelets to be non-dyadic, this introduces many additional parameters in the SC statistics. We find very efficient and effective to solve this problem by adding instead a constraint directly on the angular power spectra of the maps (see Section \ref{sec:augmentation_algo} for further details).

\subsection{Mean-field microcanonical gradient descent}\label{sec:grad_desc}

We adopt a maximum-entropy microcanonical model as the generative framework, with approximate sampling performed via gradient descent. This model has been employed previously to address the one-shot and few-shots learning problem, namely, generating new approximations of a field given only a single or a few observations \citep{Bruna2019}.

The microcanonical framework allows for efficient sampling in high dimensions: sampling proceeds by transporting high-entropy initial states to the microcanonical ensemble, which approximate the ensemble of typical samples of the process under study. 
The sampling is performed in practice by initializing a map $x$ as a random Gaussian white noise realization (whose variance matches the target homogeneous variance) and then performing a gradient descent (in pixel space in our specific case) under the loss function constraint
\begin{equation}\label{eq:micro}
\mathcal{L} = \left\|\Phi(x) - \Phi(\tilde{x})\right\|_2^2,
\end{equation}
where $\Phi$ is the latent representation of choice (in our case, SC), $x$ is the ``running'' map that is modified at each iteration of the gradient descent, $\tilde{x}$ represents the target field and  $||\cdot||_2$ is the Euclidean $\mathcal{\ell}_2$ norm. 
The sampled map is the value of $x$ at the end of the optimization.

However, the maximum-entropy microcanonical model is known to suffer from overfitting: entropy can collapse during gradient descent, causing the samples to be too similar to the target, limiting their variability. To mitigate this, one can employ mean-field gradient descent \citep{Allys:2020vld, mean_field}, which modifies the loss function (Eq.~\ref{eq:micro}) so that gradient descent acts simultaneously on a batch of $m$ maps $\{x_j\}$ with $j=1,...,m$ (initialized as independent initial white noise realizations) , moving them collectively toward the target energy:
\begin{equation}\label{eq:mean-field_micro}
\mathcal{L} = \left\|\underset{j}{\text{Ave}}\, \Phi(x_j) - \Phi(\tilde{x})\right\|_2^2, \quad \textit{(mean-field gradient descent)}
\end{equation}
where the average $\underset{j}{\text{Ave}}$ is performed over the batch of samples \(\{x_j\}\) at each iteration. 
By moving the entire batch towards the target in aggregate, mean-field gradient descent achieves a higher lower bound on entropy, thereby mitigating collapse. The final iteration of the simultaneous gradient descent gives in output a batch of $m$ samples\footnote{The mean-field microcanonical gradient descent reduces to the conventional microcanonical gradient descent if the batch size is $m=1$}. We adopt this strategy throughout the emulation procedure described below.

\subsection{Augmentation}\label{sec:augmentation_algo}

\begin{figure*}[htbp]
    \centering
        \includegraphics[clip, trim=0cm 10cm 0cm 5cm, width=1.\textwidth]{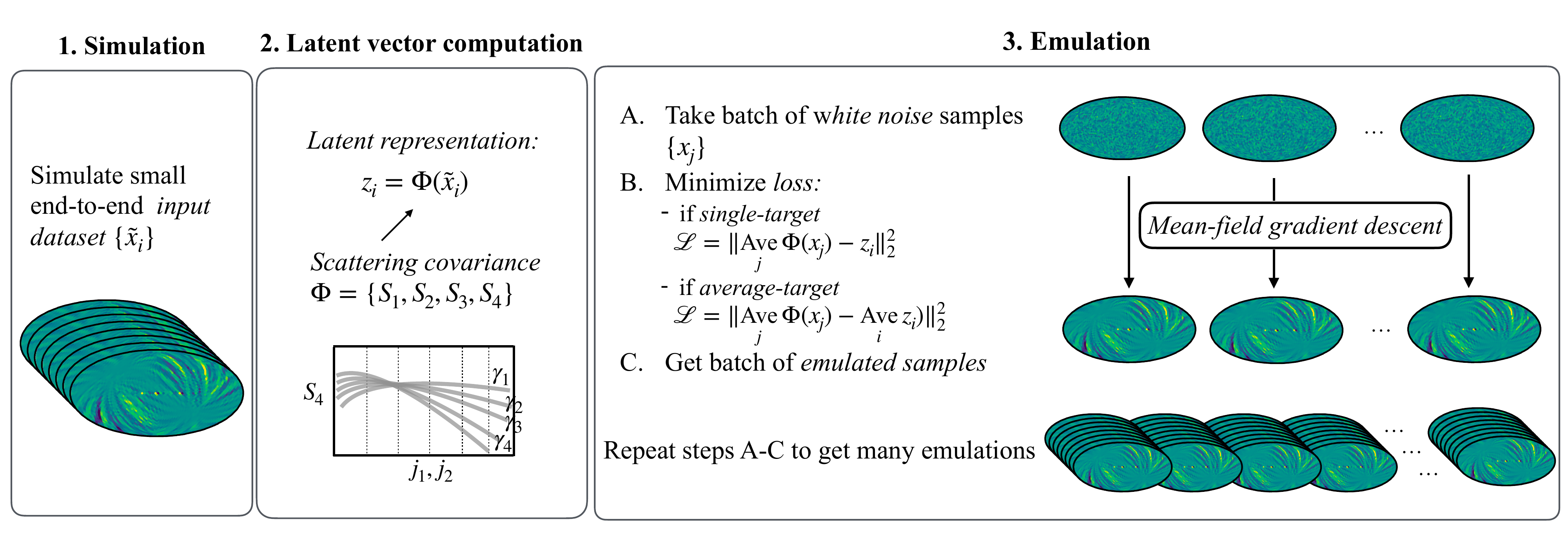}
    \caption{Schematic view of the algorithm for extreme augmentation. See Section \ref{sec:augment_summary} for details.}
    \label{fig:scheme}
\end{figure*}

We now operate under the assumption that only a limited number of end-to-end simulations is available, a common scenario not only in cosmology but also in many other fields where generating a large number of high-fidelity simulations may be prohibitively expensive. We therefore consider two representative sizes for the input dataset of simulations\footnote{We name this ``input'' dataset rather than ``training'' set, since our generative approach based on the SC does not require standard training, unlike deep-learning, for instance.}: a small dataset of 10 samples and a moderate-sized dataset of 100 maps. Naturally, we expect that the size of this dataset will affect both the quality and diversity of the generated (emulated) samples, with the larger input dataset encoding more representative information on the underlying probability distribution of the data. We compare the generative model performance in these two cases in Section~\ref{sec:results}.

\subsubsection{Single-target versus average-target}\label{sec:single_vs_average}
While Eq.~\ref{eq:mean-field_micro} defines the mean-field gradient descent loss for generating new samples from just one target field \(\tilde{x}\), our approach can leverage an ensemble of $n>1$ targets $\tilde{x}_i$ (with $i=1,...,n$).  We investigate two distinct strategies for defining the loss function incorporating multiple targets,.
 
In the first strategy, that we name average target approach, we use the average of the summary statistics from the input dataset as the target. The resulting loss function is defined as \citep{Allys:2020vld, Cheng:2023imk}
\begin{equation}\label{eq:average-target}
\mathcal{L} = \left\|\underset{j}{\text{Ave}}\, \Phi(x_j) - \underset{i}{\text{Ave}}\, \Phi(\tilde{x}_i)\right\|^2_2. \qquad\qquad \textit{(average-target)}
\end{equation}
This formulation ensures that the collective properties of the generated samples match the mean characteristics of the target ensemble.
 
The second strategy, that we name single-target approach, transfers the input probability distribution into the emulated dataset by uniformly sampling a target from the input dataset \citep{Price:2023dpj}. In standard (i.e. non-mean-field) microcanonical gradient descent, this method prescribes to randomly draw a simulation \(\tilde{x}_i\) from the input dataset and then use it as the target in the loss \ref{eq:micro}. Since we use mean-field gradient descent, we adapt this prescription as follows. First we generate a batch of $m$ new samples \(\{x_j\}\) for each of the $n$ targets in the input dataset, minimizing
\begin{equation}\label{eq:single-target}
\mathcal{L} = \left\|\underset{j}{\text{Ave}}\, \Phi(x_j) -  \Phi(\tilde{x}_i)\right\|^2_2. \qquad\qquad\qquad \textit{(single-target)}
\end{equation}
Next, we uniformly sample $N$ targets with replacement from the input dataset, and select, for each sampled target, an emulation from the corresponding batch. If the same target is sampled more than once, we ensure that each time a different emulated map is selected from the same batch. Thus, at the end, we will have a dataset of $N$ emulations.

We apply the single- and the average-target strategies to both the small and the moderate-sized input datasets, finding that the optimal approach depends on the size of the input dataset. In the following, we always show results obtained with the average-target approach for the 10-input maps case and with the single-target approach for the 100-input maps one. We discuss this topic in more detail in Section \ref{sec:sub_results}.

\subsubsection{Algorithm for extreme augmentation}\label{sec:augment_summary}

We minimize simultaneously three different losses on the SC coefficients: two for the auto-SC coefficients (Eq.\ref{eq:scat_coeff}) of the $Q$ map and $U$ map each and one on the cross-SC coefficients (i.e. computed cross-correlating the $Q$ and $U$ fields, see Appendix~\ref{sec:cross_SC}), with the latter accounting the statistical dependency between the two Stokes parameters.  

As explained in Section \ref{sec:scattering_covariance}, we complement the information encoded in the SC statistics by minimizing an additional loss on the $XX=EE, BB,EB$ angular power spectra of the maps:  
\begin{equation}\label{eq:ps_loss}
\mathcal{L} = \sum_{\ell} \left\|\frac{\underset{j}{\text{Ave}}\, C_{\ell}^{XX,(p)}(x_j)}{\underset{j}{\text{Std}}\, C_{\ell}^{XX,(p)}(x_j)} - \frac{\underset{i}{\text{Ave}}\, C_{\ell}^{XX,(p)}(\tilde{x}_i)}{\underset{i}{\text{Std}}\, C_{\ell}^{XX,(p)}(\tilde{x}_i)}\right\|^2_2,
\end{equation}
 where  
\begin{equation}
C_\ell^{XX,(p)} = \frac{1}{2\ell+1} \sum_{m=-\ell}^{\ell} |a^{XX}_{\ell m}|^p    
\end{equation}
is the angular power spectrum for the $L^p$-norm and $\text{Std}$ indicates the standard deviation of the power spectrum. Specifically, we compute this loss at each step for both the $L^1$- and the $L^2$-norm power spectra and then sum both to the SC losses.

We illustrate the method for extreme augmentation in Figure~\ref{fig:scheme}.
The augmentation process consists of three steps, with Steps 2 and 3 varying according to whether the average-target or single-target approach is chosen:
\begin{enumerate}

\item Simulation: 
\begin{itemize}
\item Generate a small set of end-to-end simulations, forming the input dataset. In our specific example, these are \texttt{HEALPix} maps including only instrumental systematics (described in detail in Section~\ref{sec:simulations}).     
\item Normalize each map by subtracting the mean and dividing by the standard deviation of the input dataset to facilitate the optimization in Step 3 below, and revert this operation upon completion.
\end{itemize}

\item Latent vector computation: 
\begin{itemize}
    \item Compute the latent representation $\Phi(\tilde{x}_i)$ of each \(\tilde{x}_i\) in the input dataset. In our specific case, $\Phi$ is the SC statistics (see Section~\ref{sec:scattering_covariance}). 
\end{itemize}

\item Emulation (i.e. sampling via gradient descent): 
\begin{itemize}
    \item  Initialize a batch of $m$ maps \(\{x_j\}\) to random Gaussian white noise realizations.
    \item At each iteration, compute $\underset{j}{\text{Ave}}\, \Phi(x_j)$ and minimize in pixel space, simultaneously for all samples in the batch, the mean-field microcanonical loss in Eq.~\ref{eq:mean-field_micro}, where the target is either the latent representation $\Phi(\tilde{x}_i)$ of one of the input maps (if we choose the single-target approach, Eq.~\ref{eq:single-target}) or the average latent vector $\underset{i}{\text{Ave}}\, \Phi(\tilde{x}_i)$ (i.e. the mean SC coefficients of all input maps, Eq.~\ref{eq:average-target}) for the average-target approach.
    \item We minimize the sum of four different losses: the auto and cross SC coefficients of the $Q$ map and $U$ maps and the loss on the $EE$, $BB$ and $EB$ angular power spectra of these maps (Eq.~\ref{eq:ps_loss}).   
    \item At the end of optimization, the resulting batch of $m$ emulated maps will have latent representations close to their respective targets.
    \item Under the single-target approach, an additional step is required: uniformly sample \(N\) targets with replacement from the input dataset and, for each target, select one emulation from its corresponding batch. 
\end{itemize}
     
\end{enumerate}
Step 3 can be repeated as many times as necessary to produce an arbitrarily large number of emulated maps based on the input dataset.

While in standard microcanonical gradient descent running the optimization for too many steps can often cause entropy collapse, here mean-field gradient descent solves this issue. The optimization reaches convergence (i.e. a plateau in the loss) in all cases after few hundred steps. We run conservatively the minimization for at least 1000 steps, which is sufficient for the global loss to decrease by 4-5 orders of magnitude, resulting in subpercent accuracy on the SC coefficients.

\section{Software}\label{sec:benchmark}

\subsection{\texttt{HealpixML} and \texttt{CMBSCAT} packages}\label{sec:software}

We compute scattering covariance coefficients on the sphere using the publicly available \texttt{HealpixML} software\footnote{\href{https://github.com/jmdelouis/HealpixML}{https://github.com/jmdelouis/HealpixML}}, written in \texttt{Python} and \texttt{TensorFlow}. 
\texttt{HealpixML} implements -- among other useful features, such as gradient descent computation and angular power spectrum implementation in \texttt{TensorFlow}-- efficient pixel-space wavelet convolutions on the \texttt{HEALPix} grid using single or multiple GPUs (but can run as well on tradition CPUs), thus enabling state-of-the-art performance for scattering analyses in spherical geometry. It supports SC calculation on 1D, 2D planar and on the \texttt{HEALPix} sphere and previous iterations of the code\footnote{Specifically, implementing the WST (see Section~\ref{sec:intro}).} have been applied to \textit{Planck} data for component separation purposes \citep{Delouis:2022yyt}. Benchmarks of \texttt{HealpixML}’s performance for this work are provided in Section~\ref{sec:sub_benchmark}.
We note that several \texttt{HealpixML} features may be broadly useful to scientists in various fields. For instance, the fast and efficient real-space convolution routines can be repurposed in other machine learning applications, while the \texttt{TensorFlow}-based implementation of the \texttt{HEALPix} angular power spectrum computation routine (i.e. \texttt{anafast}) can be incorporated into fast likelihood analyses written in \texttt{TensorFlow}.

In addition, we provide the \texttt{CMBSCAT} package\footnote{\href{https://github.com/pcampeti/CMBSCAT/}{https://github.com/pcampeti/CMBSCAT/}}, a pip-installable and user-friendly \texttt{Python}/\texttt{TensorFlow} implementation of the augmentation algorithm presented in this paper for emulating and dataset augmentation for CMB systematics. \texttt{CMBSCAT} heavily relies on the fast, efficient routines in \texttt{HealpixML} and can be readily adapted for various CMB systematics simulations or related applications. We also provide a tutorial \texttt{jupyter} notebook explaining all the features of the \texttt{CMBSCAT} package\footnote{\href{https://github.com/pcampeti/CMBSCAT/blob/main/notebook/CMBSCAT_demo.ipynb}{A \texttt{jupyter} notebook demo for \texttt{CMBSCAT} is available here.}}.  

The optimization step described in Section \ref{sec:augmentation_algo}  is performed using the L-BFGS algorithm \citep{doi:10.1137/0916069}.

 \subsection{Computational benchmarks}\label{sec:sub_benchmark}

\begin{table}[]
\caption{Computational cost for simultaneous emulation on a batch of 10 samples (each sample being made of a $Q$ and a $U$ polarization map) as a function of map resolution $N_{\rm side}$.}
  \centering
  \begin{tabular}{ccc}
    \toprule \toprule
    \(N_{\rm side}\) & Memory [GB] & GPU hrs \\
    \midrule 
    16    &   0.419  & 0.12 \\
    32    &   1.512  &  0.25 \\
    64    & 6.151  &  0.46 \\
    128   & 25.93  & 1.11 \\
    256   &  95.96   & 2.17 \\
    512   &  376.7  & 4.5 \\
    1024  &  1478  & 9.3 \\
    2048  & 5804  & 19.3 \\
    \bottomrule \bottomrule
  \end{tabular}
    \tablefoot{The table shows the peak memory usage in GB and speed in GPU hours for each resolution level, obtained using double precision floats on an NVIDIA A100 64 GB GPU. Values up to $N_{\rm side}=128$ are measured, while the ones at higher resolution are obtained extrapolating the trend of the measured points (see Section~\ref{sec:sub_benchmark} for details).}
  \label{tab:comp_cost}
\end{table}

The computational expense of TOD simulations can vary significantly depending on the instrument characteristics and the desired level of detail. 
In general, the cost of TOD simulations for CMB experiments is largely independent of the final map resolution. Instead it is predominantly driven by the complexity of the systematics effects and noise included in the timeline and by the assumptions made in the calibration and map-making steps, such as the correlation length of the noise assumed in the map-making \citep{map_making_1, map_making_2}.
In many past efforts, massive Monte Carlo simulation campaigns have consumed  $\mathcal{O}(100)$ million CPU hrs to produce only $\mathcal{O}(10^3)$ high resolution ($N_{\rm side}=2048$) simulations \citep[see e.g. ][]{Planck:2015txa}, amounting roughly to $\mathcal{O}(10^5)$ CPU hrs per map.

The augmentation approach advocated here bypasses additional TOD generation and the subsequent map-making steps\footnote{We note that also the computational cost of map-making might be relevant if not comparable with the one of TOD production. The approach proposed in this paper can in principle avoid also further costs associated to this step.} by directly emulating at map-level, once the generative model has been built on a small ensemble of simulations. In contrast to TOD simulations, the emulator offers the possibility of directly generating low-resolution maps where appropriate, resulting in significant saving in computational resources. Its computational cost and memory requirements scale linearly with the number of orientations $R$ and number of wavelet scales $J$ employed in the analysis. Table~\ref{tab:comp_cost} summarizes the computational and memory costs of the emulator at several \(N_{\rm side}\) values on a single NVIDIA A100 64 GB GPU, assuming the gradient descent is performed simultaneously on a batch of 10 maps for 1000 steps on the four losses described in Section~\ref{sec:augment_summary}, using double precision floats\footnote{Using single precision would imply approximately a reduction of a factor 2 in GPU hrs and memory requirements. Also, using short floats would reduce the requirements shown in Table \ref{tab:comp_cost} by a factor 4 and might be a viable alternative in some cases.}. Values in this table for resolution up to $N_{\rm side}=128$ are measured, and then extrapolated to higher resolution fitting a power-law of the kind $x(N_{\rm side}) = A*(N_{\rm side})^{a}$ (where $x$ is either the cost in GPU hours or the memory in GB, $A$ and $a$ are constants estimated during the fit) to the measured data. The estimated execution time (in GPU hrs) to generate 10 new emulated maps scales roughly linearly with $N_{\rm side}$ and is significantly smaller that the respective cost of a TOD simulation in CPU hrs, even at very high resolution (e.g. compare 1.93 GPU hrs per map at $N_{\rm side}=2048$ versus $\mathcal{O}(10^5)$ CPU hrs for TOD).  While emulation of high-resolution maps (e.g., \(N_{\rm side}=2048\)) might still incur a significant memory cost, most practical applications—especially those focusing on \(B\)-mode polarization detection in full-sky experiments—do not require such high resolution. The computational cost is dominated by the wavelet convolution at each iteration during gradient descent, which requires a large graph in memory to be computed efficiently. For high-resolution cases, a multi-scale approach \citep[e.g.][]{Marchand:2022fxp}, in which large scales are computed first and small scales are computed locally, should be considered instead, reducing significantly the memory footprint below what is shown in Table~\ref{tab:comp_cost} for large $N_{\rm side}$ values. We leave further exploration of this point to future work.

Finally, our current implementation leverages the MPI-parallelized design of \texttt{HealpixML} to efficiently scale on multiple CPUs or GPUs. Although the primary computational bottleneck remains the wavelet convolution in real space on the sphere (which is considerably faster when performed instead on a small flat-sky patch) at each iteration during gradient descent, we expect that continued development and optimization—guided by the requirements on augmentation on noise and systematics templates from collaborations such as \textit{LiteBIRD}, SO and CMB-S4—will further enhance the efficiency of our simulation framework.

\section{Validation}\label{sec:results} 

\begin{figure*}[htbp]
\centerline{\includegraphics[clip, trim=0.5cm 1.5cm 0.5cm 1cm, width=1.1\linewidth]{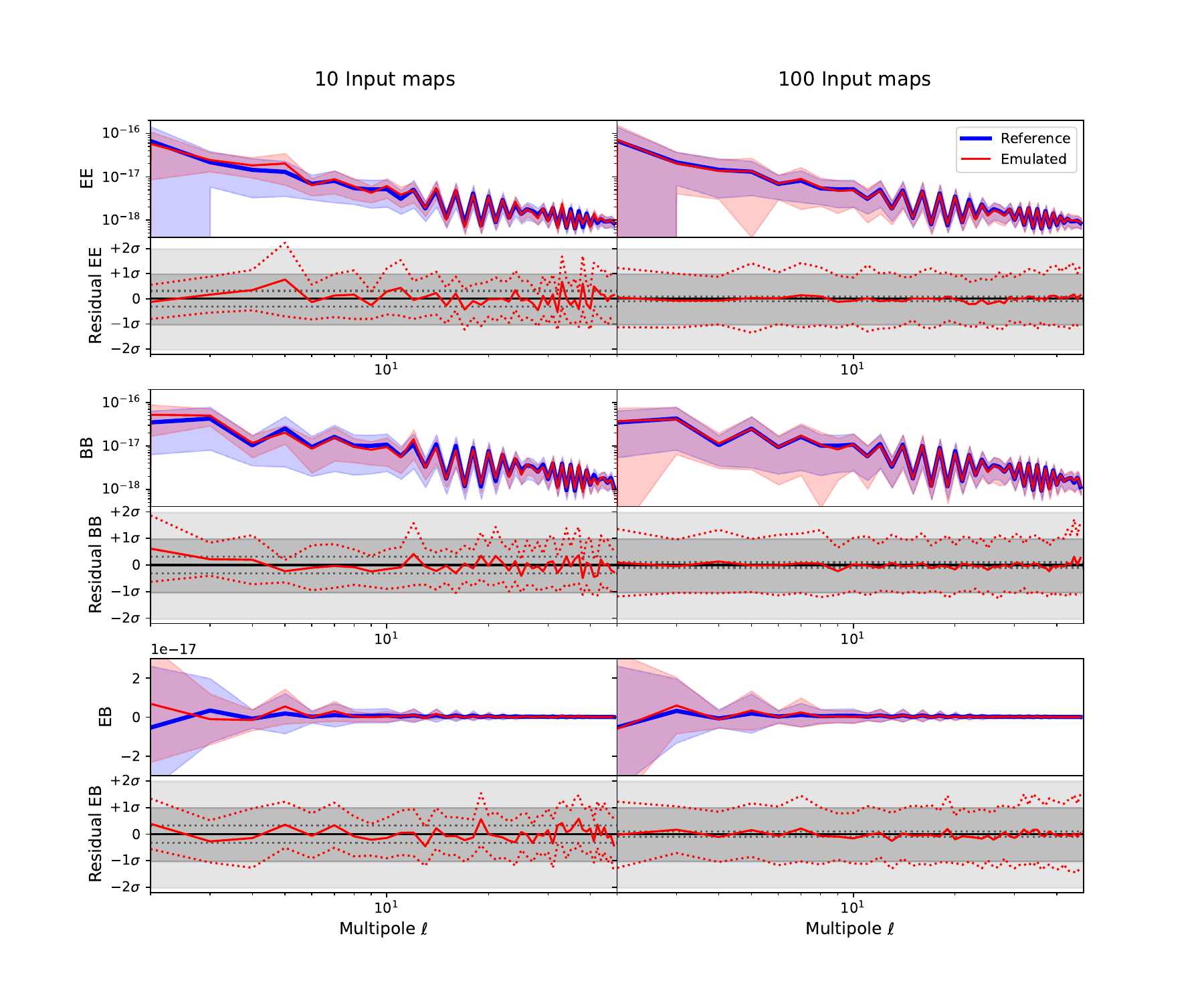}}
\caption{Comparison of the \(EE\), \(BB\), and \(EB\) angular power spectra for the emulated maps (red) versus the reference dataset (blue), shown for the $N_{\rm input} = 10$ (left) and $N_{\rm input} = 100$ (right) cases. In each row, the top panel displays the average power spectra with 1-$\sigma$ shading, while the bottom panel shows the residual, 
$\Delta_\ell = (\langle C_\ell^\mathrm{emu}\rangle - \langle C_\ell^\mathrm{ref}\rangle)/\sigma_\ell^\mathrm{ref}$ (solid red) along with $\Delta_\ell^\pm = (\langle C_\ell^\mathrm{emu}\rangle - \langle C_\ell^\mathrm{ref}\rangle \pm \sigma_\ell^\mathrm{emu})/\sigma_\ell^\mathrm{ref}$ (dotted red), where $\sigma_\ell^\mathrm{ref}$ is the standard deviation from the reference set and $\sigma_\ell^\mathrm{emu}$ is the emulated dataset's one. In both the 10-map and 100-map cases, the mean spectra and variance per multipole closely match those of the reference set, and the residuals lie well within the 1-$\sigma$ band. The bottom panels also show in dotted gray the standard error on the mean for a sample of size $N_{\rm input}$, that is $\pm \sigma_{\rm ref}/\sqrt{N_{\rm input}}$, with $N_{\rm input}=10$ ($100$) in the left (right) plots. Notably, the scatter of the mean residual is roughly within this band in both cases, indicating success of the augmentation process. See Section \ref{sec:sub_results} for further details.}
\label{fig:validation}
\end{figure*}

\begin{figure*}[ht]
\centering
\includegraphics[width=\textwidth]{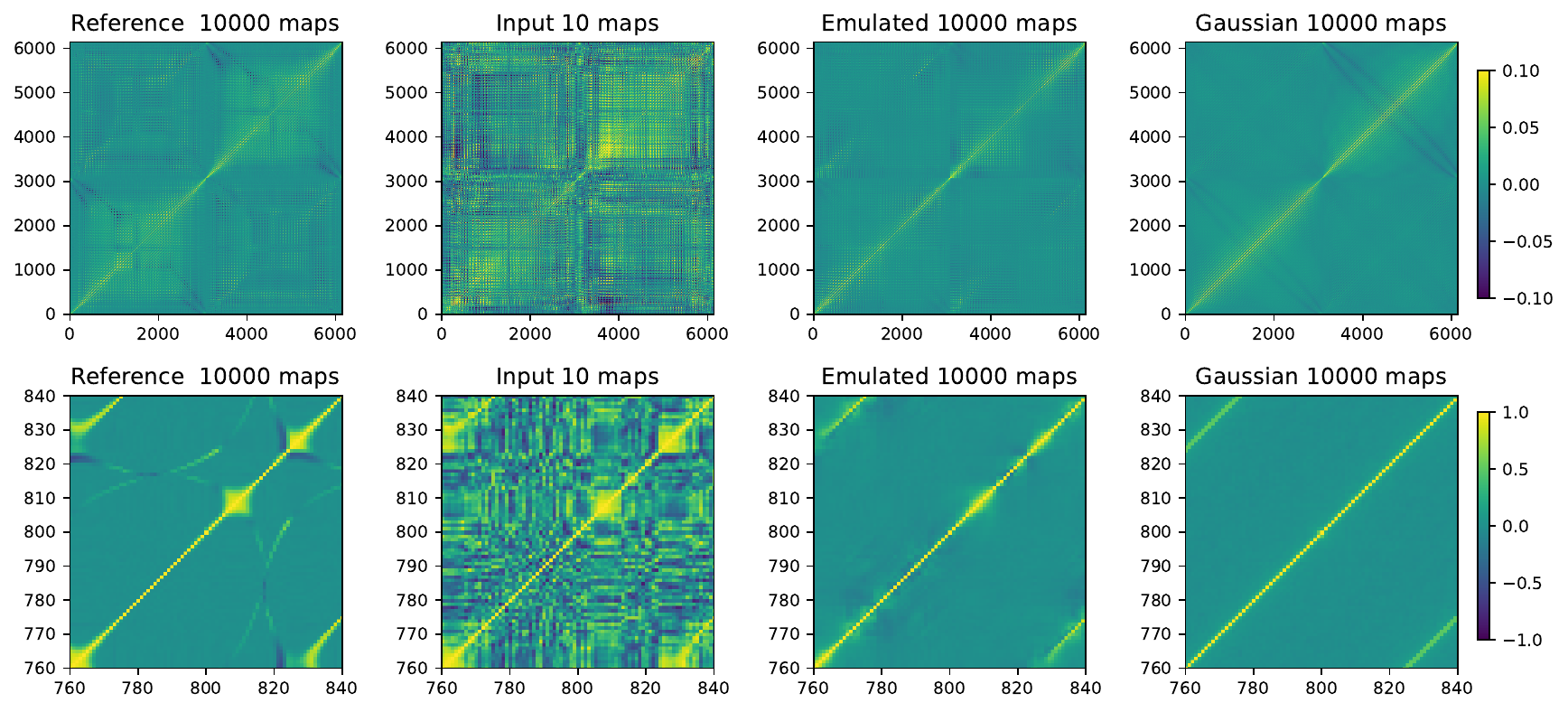}
\caption{Upper panel: Empirical pixel correlation matrices estimated from the reference dataset, the input dataset, the emulated dataset and the Gaussian dataset in the $N_{\rm input}=10$ case. The color-scale runs between -0.1 and 0.1 to highlight the structures and allow for visual comparison. See Section \ref{sec:results} for details. Bottom panels: Zoom-in between pixel 760 - 840 of the corresponding covariance matrix in the upper panel. Note that here the color-scale runs between -1 and 1. See Section \ref{sec:sub_results} for further details.}
\label{fig:covmats}
\end{figure*}

\begin{figure*}[ht]
\centering
\includegraphics[width=\textwidth]{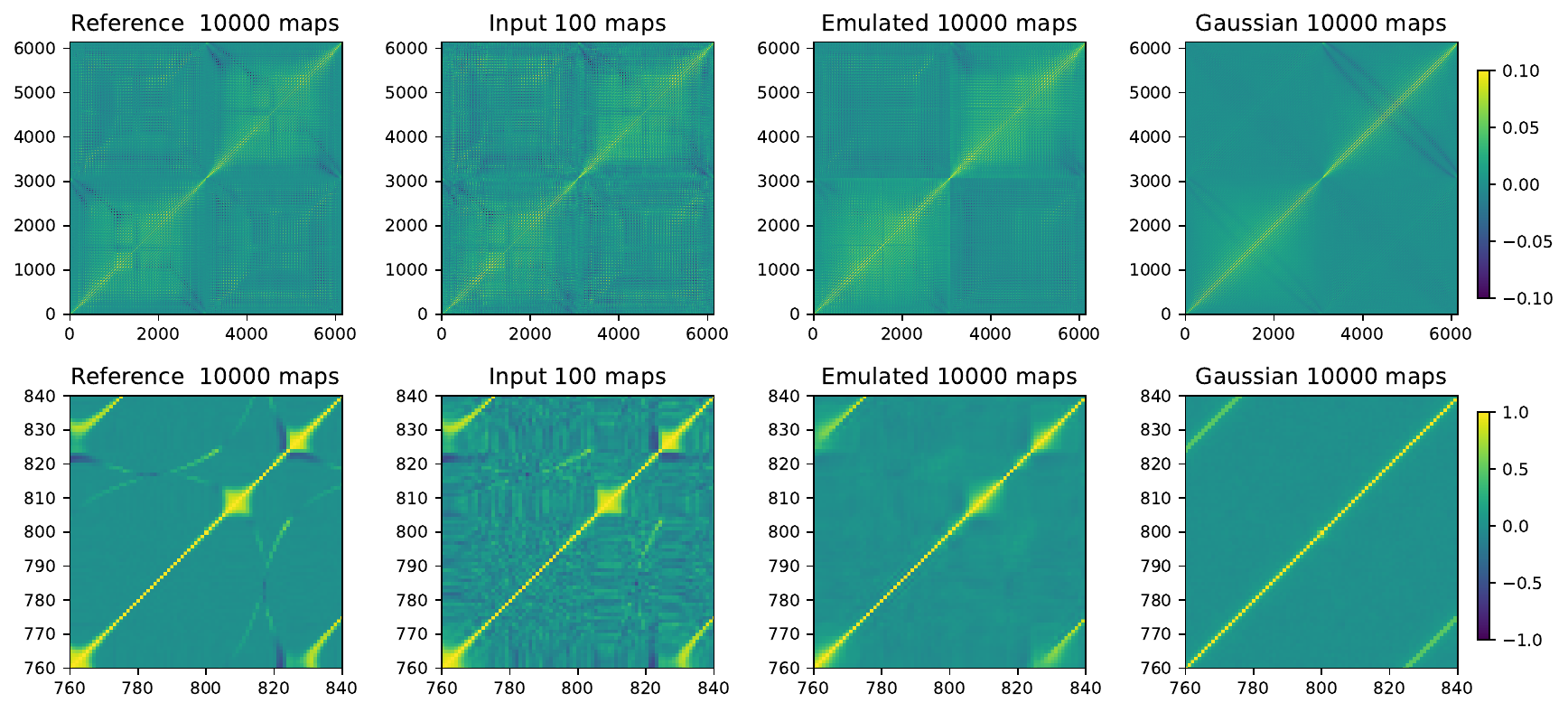}
\caption{Same as Figure \ref{fig:covmats} but for the $N_{\rm input} = 100$ case. See Section \ref{sec:sub_results} for further details.}
\label{fig:covmats_100maps}
\end{figure*}

\begin{figure}[htbp]
\centerline{\includegraphics[width=1.\linewidth]{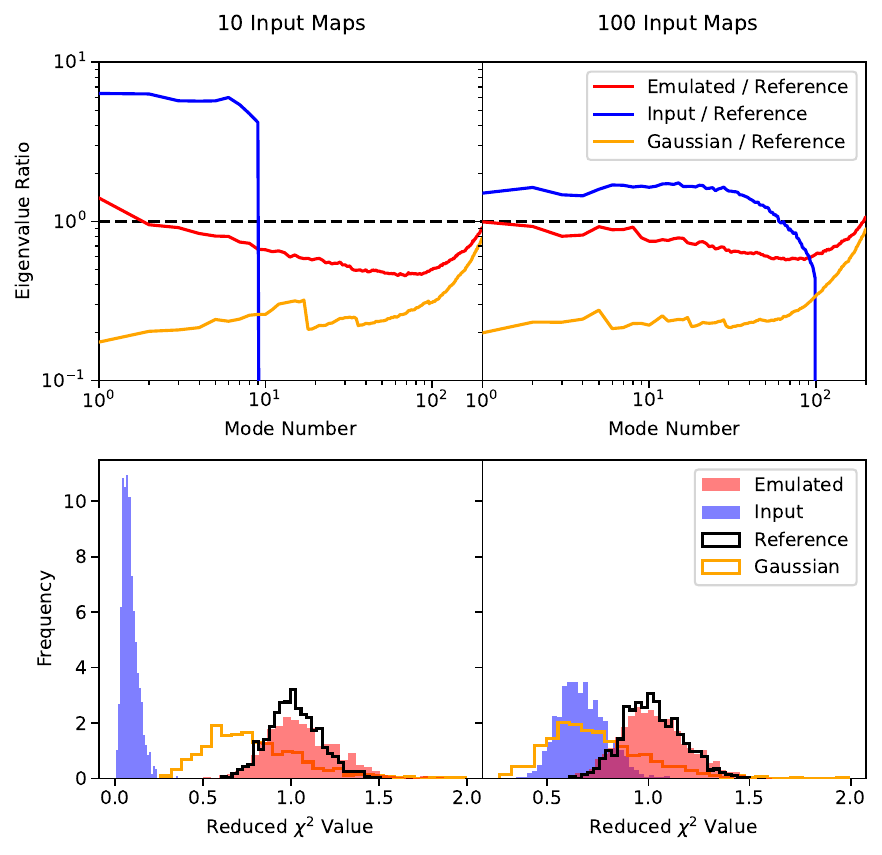}}
\caption{Top panel: Ratio of eigenvalues of the input, emulated and Gaussian pixel covariance matrices in Fig.\ref{fig:covmats} to the reference covariance's ones. Bottom panel: Histograms of chi-square values computed using 1000 maps from the reference dataset and the four different covariance matrices. See Section \ref{sec:sub_results} for details.}
\label{fig:eigen}
\end{figure}

In this section we present the visual and statistical validation of the generative model. First, we define the various datasets used in the validation strategy. Next, we introduce the statistical tests employed to compare the model outputs with the target distribution. Finally, we discuss the results of the validation campaign.

\subsection{Datasets definition}\label{sec:datasets}

The validation framework relies on seven different datasets, each playing a distinct role in testing different aspects of the generative model. We define them in the following.

\paragraph{Reference dataset.} We begin with a set of 10,000 end-to-end TOD simulations (as described in Section \ref{sec:simulations}), from which we extract 10,000 corresponding $Q$ and $U$ maps. We call this the original dataset in the following.  We use the original dataset exclusively to estimate the full $QU$ pixel-to-pixel covariance and pixel-to-pixel correlation matrices, hereafter referred to as the reference covariance and correlation matrices. These matrices serve as the ground truth for the covariance (and correlation) structure we aim to reconstruct. We then generate a reference dataset of 10,000 maps as random Gaussian samples from this reference covariance. Given that these realizations are sampled from the reference covariance, we can ignore the sample variance introduced by estimating an empirical $6144\times6144$ covariance matrix from only 10,000 simulations and treat the reference covariance as the target covariance that the emulator should ideally reproduce (i.e. the ground-truth). This choice is further justified by the fact that the reference covariance matrix appears to encode quite well the pixel-to-pixel spatial information in the original dataset: this can be understood considering that we simulated the original dataset as an additive random Gaussian fluctuation in the TOD. We do not expect therefore significant differences in our results when emulating the original dataset or the reference dataset. We further detail the reasons behind this choice in Appendix~\ref{sec:appendix_Gaussian}. 

The reference dataset provides a basis for comparing various statistical properties and ensuring that the emulator captures the true underlying distribution described by the reference covariance.

We emphasize that generating the reference dataset as Gaussian realizations from the pixel-to-pixel reference covariance matrix results in an inhomogeneous Gaussian field that cannot be fully characterized by its power spectrum alone, since the latter lacks any spatial information related to the specific information at each pixel. The additional, ``beyond power spectrum'' information encoded in the SC thus plays a crucial role in constraining the realizations.

\paragraph{Input datasets.} To mimic realistic scenarios where only a limited number of simulations are available, we generate two input datasets by drawing random Gaussian realizations from the reference covariance: 
\begin{enumerate} 

\item A small input dataset containing $N_{\rm input} = 10$ realizations. 

\item A moderate size input dataset containing $N_{\rm input} = 100$ realizations. 

\end{enumerate} 
As for the reference dataset, since also these realizations are sampled from the reference covariance, the latter represents the ground-truth covariance that the emulator should aim at reproducing. For each input dataset, we compute the empirical pixel covariance (and correlation) matrices (input covariance (correlation) matrix).  Since the number of simulations used to estimate them is lower than the number of their degrees of freedom, the input covariances will be plagued by high sample variance and will be rank-deficient (i.e. under-determined), leading to instability under inversion \citep{Hartlap:2006kj}.

\paragraph{Emulated datasets.} We apply the generative model to each of the two input datasets, following the average-target strategy for the $N_{\rm input}=10$ set and the single-target one for $N_{\rm input}=100$ (see Section~\ref{sec:single_vs_average}) and augmenting each to create two corresponding emulated datasets of 10,000 maps each. From these maps, we compute the pixel covariance and correlation matrices—hereafter referred to as the emulated covariance (correlation) matrices—to evaluate how accurately the emulator reconstructs the target covariance.

\paragraph{Gaussian datasets.} For additional comparison, we also construct two Gaussian datasets using a naive augmentation strategy. Specifically, from the isotropic angular power spectra of each map in the 10-map and 100-map input datasets\footnote{We find completely similar behavior of the Gaussian correlation matrices as the ones shown in Figs.~\ref{fig:covmats} and Figs.~\ref{fig:covmats_100maps} in Section~\ref{sec:results} when generating maps from the average power spectrum of the input dataset instead.}, we generate 1,000 and 100 Gaussian random realizations, respectively, leading to two datasets of 10,000 maps each. The corresponding covariance matrix is denoted as the Gaussian covariance.

In summary, the validation involves seven different datasets: a reference dataset, two input sets, two emulated sets and the two Gaussian datasets.

\subsection{Visual and statistical tests}\label{sec:stat_tests}

We employ visual and statistical tests to verify that the generated maps are consistent with the true underlying distribution.

\begin{enumerate}

\item Visual inspection. We compare visually both emulated map samples (Fig.~\ref{fig:maps}) and emulated pixel covariances (Figs.~\ref{fig:covmats} and \ref{fig:covmats_100maps}) with the reference ones.

\item Angular power spectra.
We compare the $EE$, $BB$, and $EB$ power spectra of the emulated maps against those computed from the reference dataset (see Fig.~\ref{fig:validation}). We expect the emulator to be able to reproduce both the average power spectrum and the correct variance-per-multipole.

\item Pixel covariance matrices.  
We visually and quantitatively compare the empirical pixel correlation matrices of the emulated datasets with the reference one (Figs.~\ref{fig:covmats} and \ref{fig:covmats_100maps}). In particular, we examine the ratio of the eigenvalues of the covariance matrices (Fig.~\ref{fig:eigen}) and perform a $\chi^2$ test (see point 4 below) to assess the fidelity of the off-diagonal structure.

\item Reduced $\chi^2$ statistics.  
We compute the reduced $\chi^2$ statistic for 1000 maps in the reference set using each covariance matrix:
\begin{equation}\label{eq:chi-square}
    \chi^2 = \frac{\mathbf{m}^T \, \mathbf{C}^{-1} \, \mathbf{m}}{d}\,,
\end{equation}
where $\mathbf{m}$ represents a map from the reference dataset, $\mathbf{C}$ is one of the covariance matrices (reference, input, emulated, or Gaussian), and $d$ is the number of degrees of freedom. In practice, we perform a principal component analysis (PCA) on $\mathbf{C}$ and retain the largest 150 eigenvalues of the reference, emulated and gaussian covariances (accounting for approximately 90\% of the total variance) and the largest 10 or 100 eigenvalues of the input covariance (depending whether $N_{\rm input}=10$ or $100$), when computing $\chi^2$. The histogram of these reduced $\chi^2$ values (see Fig.~\ref{fig:eigen}) provides another quantitative measure of statistical consistency.
\end{enumerate}

We apply all of these tests to both the small (10 maps) and moderate (100 maps) input datasets. Additional validation metrics, such as SC coefficients, Minkowski functionals and probability distribution functions (PDFs) of the maps are presented in Appendix \ref{sec:extra_validation}.

 \subsection{Results}\label{sec:sub_results}
Results shown in this section are obtained using the average-target strategy for the $N_{\rm input}=10$ case and the single-target strategy for the $N_{\rm input}=100$ one (see Section~\ref{sec:single_vs_average}).
We find that the optimal choice of the target for gradient descent depends on the size of the input dataset. When only a few tens (or fewer) of simulations are available, using the average-target approach (Eq.~\ref{eq:average-target}) yields more stable results compared to the single-target strategy. This is because, in our case, a small sample is sufficient to capture somewhat accurately the ``true'' average of the SC coefficients across the reference dataset, leading to visually improved empirical covariance matrices and enhanced agreement in both covariance eigenvalues and \(\chi^2\) values with the reference. Also for datasets comprising hundreds of simulations, using the average-target approach allows for improvements across all metrics described in Section~\ref{sec:stat_tests} when compared to the same quantities evaluated without any augmentation. However, in this case the single-target method (Eq.~\ref{eq:single-target}) allows for further improvements over the average-target results. This is because the single-target strategy more effectively transfers the full statistical distribution encoded in the input dataset to the emulated outputs, capturing true structures in the covariance matrix beyond the capabilities of the average-target.

The results of the validation are organized into four main parts, following Section~\ref{sec:stat_tests}: visual inspection of the maps, power spectrum consistency, pixel-to-pixel covariance matrix reconstruction and reduced $\chi^2$-test in pixel space.

\paragraph{Visual inspection.}
Figure~\ref{fig:maps} shows a side-by-side comparison of three input maps and three corresponding emulated samples synthesized from each of the inputs. Specifically, we show in this figure the emulated maps obtained from the moderate-sized input dataset. For each row, the leftmost column displays the input (target) map from this dataset, while the other columns present distinct emulated samples obtained from the specific target map. We get very similar visual comparison when emulating instead from the small (10-maps) input set, which we do not show here.
Notably, the emulated maps capture key features of the target, such as the scanning strategy stripes and the ecliptic pole patterns, at the right positions while also displaying clear differences, similar to those between the different input maps. This suggests that the emulator successfully balances predictiveness (accurate feature reconstruction) with representativeness (maintaining sample variability).

\paragraph{Angular power spectra.}

In Fig.~\ref{fig:validation}, we compare the \(EE\), \(BB\), and \(EB\) angular power spectra of the emulated maps to those from the reference dataset. To quantify their difference, we define the residual
\[
\Delta_\ell \;=\; \frac{\langle C_\ell^\mathrm{emu}\rangle \;-\; \langle C_\ell^\mathrm{ref}\rangle}{\sigma_\ell^\mathrm{ref}},
\]  
where \(\langle C_\ell^\mathrm{emu}\rangle\) and \(\langle C_\ell^\mathrm{ref}\rangle\) are the average power spectra in the emulated and reference sets, respectively, and \(\sigma_\ell^\mathrm{ref}\) is the standard deviation measured in the reference set. As shown in the bottom panels (solid red lines), \(\Delta_\ell\) remains well within the 1\(\sigma\) band across the multipoles.
To compare the variance-per-multipole in the two datasets, we also plot (as dotted red lines) 
\[
\Delta_\ell^\pm \;=\; \frac{\langle C_\ell^\mathrm{emu}\rangle \;-\; \langle C_\ell^\mathrm{ref}\rangle \;\pm\; \sigma_\ell^\mathrm{emu}}{\sigma_\ell^\mathrm{ref}},
\]  
where \(\sigma_\ell^\mathrm{emu}\) is the standard deviation in the emulated dataset. Both the small and moderate input cases agree very well with the reference set in terms of mean spectra and variance. Notably, even when trained on only 10 input maps, the emulator still reproduces the mean and variance of the angular power spectrum with good fidelity. We also plot in the bottom panels of Fig.~\ref{fig:validation} two dotted gray lines indicating the standard error on the mean for a sample of size $N_{\rm input}$, namely $\pm \sigma_{\rm ref}/\sqrt{N_{\rm input}}$, with $N_{\rm input}=10$ in the left plots and $N_{\rm input}=100$ in the right plots. Notably, the scatter on the mean residual is roughly within this band in both cases, indicating that the augmentation process is approximately at least as good as having $N_{\rm input}$ new direct draws from the reference dataset: the emulated dataset reproduces the mean power spectra as closely as one would expect if one had roughly $N_{\rm input}$ truly independent draws from the true distribution, without adding any significant extra bias beyond that expected from the small input dataset.

\paragraph{Pixel covariance matrices.}
Figures \ref{fig:covmats} and \ref{fig:covmats_100maps} show the empirical pixel correlation matrices for the reference, input, and emulated datasets\footnote{We show correlation matrices obtained from maps in \texttt{HEALPix} RING ordering.}. In the 10-map case, the input correlation is visibly noisy, reflecting the small sample size. However, the emulated correlation matrix—computed from 10,000 generated maps—exhibits a much cleaner structure while retaining some of the main features of the reference correlation matrix. With 100 maps, the input correlation matrix suffers less from sample noise, but the emulated one continues to capture both prominent and more subtle features seen in the reference significantly better than the input one. We note that some of the more subtle ``filamentary'' structures in reference covariance in the bottom panel of Fig.~\ref{fig:covmats_100maps} appear slightly smoothed over more pixels in the emulated one. This is connected to the number of wavelet orientations $R$ and the size of the wavelet convolution kernel (i.e. $3\times3$) we chose (see Section~\ref{sec:scattering_covariance}). While this specific choice gives already satisfactory results in all key metrics considered, we note that it is sufficient to increase both of these parameters to recover these subtle off-diagonal features more sharply.

\paragraph{Eigenvalues and reduced $\chi^2$ statistics.} To quantify these observations, we compare in Fig.\,\ref{fig:eigen} the eigenvalues of the different covariance matrices and the reduced \(\chi^2\) values for 1000 maps in the reference set using each covariance matrix (Eq.~\ref{eq:chi-square}). These comparisons again indicate that the emulated maps help more accurately reproduce the true underlying distribution compared to the naive estimation from 10 or 100 input maps. 
Because the generative model can produce new (approximate) independent realizations of the underlying stochastic process, augmenting the input dataset reduces the sample variance and, consequently, the bias it introduces in the inverse covariance matrices \citep{Hartlap:2006kj}. Therefore, the eigenvalues and the reduced \(\chi^2\) distribution derived from the emulated covariance matrix match more closely those from the reference covariance, surpassing the performance of the covariance naively estimated only from input simulations, in both the $N_{\rm input} = 10$ and the $N_{\rm input} = 100$ cases. In particular, the top panels in Fig.\,\ref{fig:eigen} shows that the generative model-driven augmentation helps recover the first \(\sim 150\) eigenvalues significantly better than the naive estimation, while the bottom panels illustrate that the reduced \(\chi^2\) values from the emulated covariance cluster around 1, and have a distribution consistent with the reference. In contrast, the reduced \(\chi^2\) values from the input covariance cluster are consistently off from 1, indicating a very poor estimation of uncertainties in the reference set.

In addition to the emulation-based approach, we also consider a naive augmentation strategy—labeled “Gaussian” in Figs.~\ref{fig:covmats}–\ref{fig:eigen}—which consists in generating additional maps by sampling new Gaussian random realizations from the isotropic power spectrum of each input map\footnote{We find completely similar results when generating maps from the average power spectrum of the input dataset instead.}. Although this approach can produce visually “cleaner” off-diagonal structures in the covariances (see the rightmost panels of Figs.~\ref{fig:covmats} and \ref{fig:covmats_100maps}), it fails to replicate even the brighter, more pronounced features observed in the reference correlation matrix. Comparisons of both eigenvalues and the reduced \(\chi^2\) histograms (Fig.~\ref{fig:eigen}) confirm that this naive augmentation method falls short of capturing the full covariance structure and performs significantly worse than the proposed emulation-based strategy.

\section{Conclusions}\label{sec:conclusions}

In this work, we introduced a novel emulator for CMB systematics datasets based on the scattering covariance statistics. Unlike neural-network models developed for similar purposes \citep[e.g., diffusion models,][]{diffusion}, this approach is fully interpretable and requires only a small number of high-fidelity simulations to generate new accurate and statistically consistent samples. We demonstrate that a small ensemble can be effectively augmented with such emulator to achieve robust statistical performance in several key tasks of cosmological analyses. We deploy a number of tests in order to demonstrate this statistical robustness, among them the mean and variance of power spectrum estimates and the reduction in sample variance and its associated bias in inverse pixel covariance matrices.

The emulator is validated using realistic, highly inhomogeneous map simulations of CMB systematics, which are especially challenging to emulate. We tested the emulator on both small (10 input maps) and moderate-sized (100 input maps) datasets. As expected, increasing the number of input simulations improves the quality of emulation. Nonetheless, even when trained on just 10 simulations, the emulator produces remarkable results in both visual and statistical tests.

Although generative models based on the SC statistics naturally targets homogeneous data, we nonetheless achieve successful emulation in this more complex context, characterized by strongly inhomogeneous statistics. Consequently, we are confident that the SC approach -- especially when complemented by the orientation-informed wavelets that we introduced here -- can be applied to a broad range of observational scenarios featuring similar large-scale instrumental structures. 

The framework can be adapted also to multiplicative or signal-dependent systematics, beyond the additive noise example studied here. Instead of performing data augmentation by sampling independent realizations of the noise, we should instead recast the task as an inverse problem, describing the multiplicative systematic as the action of a stochastic forward model that maps a clean signal to a contaminated one. This can be achieved, for instance, by estimating the joint and cross scattering statistics of the signal and the contaminant, and then sampling new noise realizations conditioned on a given signal map. Multi-channel generative models with cross ST statistics have been already introduced in \citet{Regaldo23}. Also, \citet{Delouis:2022yyt} shows an example of map sampling whose cross-statistics with another signal is constrained (i.e. $Q/U$ polarized dust emission at 353 GHz conditioned with a total intensity dust emission map).

The emulator can operate entirely on GPUs, which are in this case significantly more energy-efficient than CPUs, offering substantial practical benefits. For example, augmenting a dataset of 100 low-resolution $Q$ and $U$ maps (with \(N_{\rm side}=16\)) by a factor of 100 requires roughly 120 GPU-hours on a single NVIDIA A100 64 GB GPU—a negligible cost compared to the millions of CPU-hours typically needed for full end-to-end simulations in full-sky experiments. This might enable analyses on both data and simulations that were previously unfeasible due to lack of computational resources. Moreover, the energy efficiency of GPUs further suggests that our method could significantly reduce the carbon footprint of extensive simulation campaigns.  

Looking forward, we plan to make our method efficient also for higher-resolution maps and benchmark its performance in multi-GPU environments, which are necessary to overcome current GPU memory limitations during simultaneous gradient descent. Given the strong scaling of the framework to multiple GPUs, there is a clear road-map ahead for its development and testing. On the algorithmic front, incorporating advanced techniques such as meta-optimization may further enhance the stability and accuracy of the loss minimization process. 

Moreover, our emulator opens the door to a range of applications beyond augmentation, including simulation-based inference for cosmological parameters and the denoising of complex instrumental systematics. Simulation-based inference might indeed prove necessary to fully exploit future cosmological datasets, and avoid the hard (or sometimes even impossible) task of characterizing complex systematics effects at the likelihood level \citep{Wolz:2023gql}. Also Quadratic Maximum-Likelihood (QML) power spectrum estimators and pixel-based likelihoods \citep{Tegmark:1996qs} might benefit from our approach: augmenting the datasets used to compute the empirical noise pixel covariance matrices needed by both methods might help to achieve optimal errors bars, especially in scenarios where only few hundreds simulations are available. As a concrete example, we plan to use our method to mitigate ADC non-linearities \citep{Planck:2016kqe} in Planck polarization data \citep[see e.g. ][for a simulation-based approach]{Wolz:2023gql}, aiming for more accurate estimates of the optical depth of reionization and tensor-to-scalar ratio.  

In summary, our approach advocates for a paradigm shift in simulation strategies:  high-fidelity emulation does not require an extensive number of simulations, but rather a few carefully crafted ones that accurately represent the experimental noise model and sky. Rather than producing vast numbers of lower or intermediate-quality simulations, future pipelines for cosmological surveys—such as Euclid, \textit{LiteBIRD}, SO, CMB-S4 and Rubin-LSST, should focus on generating a few high-precision simulations. These can then be efficiently augmented using the framework presented here. 

\section*{Data availability}
In the spirit of reproducibility and accessibility, we provide to the community both a general framework for computing the SC statistics \texttt{HealpixML} at \url{https://github.com/jmdelouis/HealpixML} and the emulator \texttt{CMBSCAT} at \url{https://github.com/pcampeti/CMBSCAT/} we developed.

\begin{acknowledgements} 
We thank: V. Amitrano, G. Galloni and W.L. Kimmy Wu for advice and fruitful discussions;  E. Calore and A. Miola for help in setting up the computational environment. We acknowledge the use of computing facilities provided by the INFN theory group (I.S. InDark) at CINECA and the COKA machine at INFN Sezione di Ferrara. 

This work has been supported by the Fondazione ICSC, Spoke 3 Astrophysics and Cosmos Observations. National Recovery and Resilience Plan (Piano Nazionale di Ripresa e Resilienza, PNRR) Project ID CN\_00000013 ``Italian Research Center on High-Performance Computing, Big Data and Quantum Computing''  funded by MUR Missione 4 Componente 2 Investimento 1.4: Potenziamento strutture di ricerca e creazione di ``campioni nazionali di R\&S (M4C2-19 )'' - Next Generation EU (NGEU). 

M.G., M.L. and L.P. acknowledge the financial support from the INFN InDark initiative and from the COSMOS network through the ASI (Italian Space Agency) Grants 2016-24-H.0 and 2016-24-H.1-2018. M.G. and M.L. are funded by the European Union (ERC, RELiCS, project number 101116027). M.G. acknowledges support from the PRIN (Progetti di ricerca di Rilevante Interesse Nazionale) number 2022WJ9J33.

\end{acknowledgements}

\bibliographystyle{aa}
\bibliography{biblio/PC}

\begin{appendix}

\section{Orientation-informed wavelets}\label{sec:oriented_wavel}

\subsection{Spherical wavelet definition}
The SC statistics used in this study is based on oriented wavelet convolutions on the sphere. The convolution uses the \texttt{HealpixML} software \footnote{\href{https://github.com/jmdelouis/HealpixML}{https://github.com/jmdelouis/HealpixML}} based on \texttt{HEALPix} \citep{healpix2005} to describe the pixelized data on the sphere. 
The wavelet kernel is different for each pixel as the \texttt{HEALPix} pixel could have a different shape, in particular near the pole and at large scale. Thus, the original complex kernel defined by the equation
\begin{eqnarray}
W_{j,\Theta} & = &\left(\cos(2\pi\alpha)+j \sin(2\pi\alpha)\right)e^{-(\hat{\theta}^{2}+\hat{\phi}^{2})}  \\
N_{side} &=& 2^{j} \\ \nonumber
\hat{\theta} &=&N_{side} \left(\theta-\frac{\pi}{2}\right) \nonumber\\
\hat{\phi}&=&N_{side} \phi \nonumber \\
\alpha &=& \hat{\phi}\cos{\Theta}+\hat{\theta}\sin{\Theta}  \nonumber
\end{eqnarray}
is convolved with the data to compute the wavelet of the direction $\Theta$ at the scale $j$.

\begin{figure}[]
\centering
\includegraphics[width=0.5\textwidth]{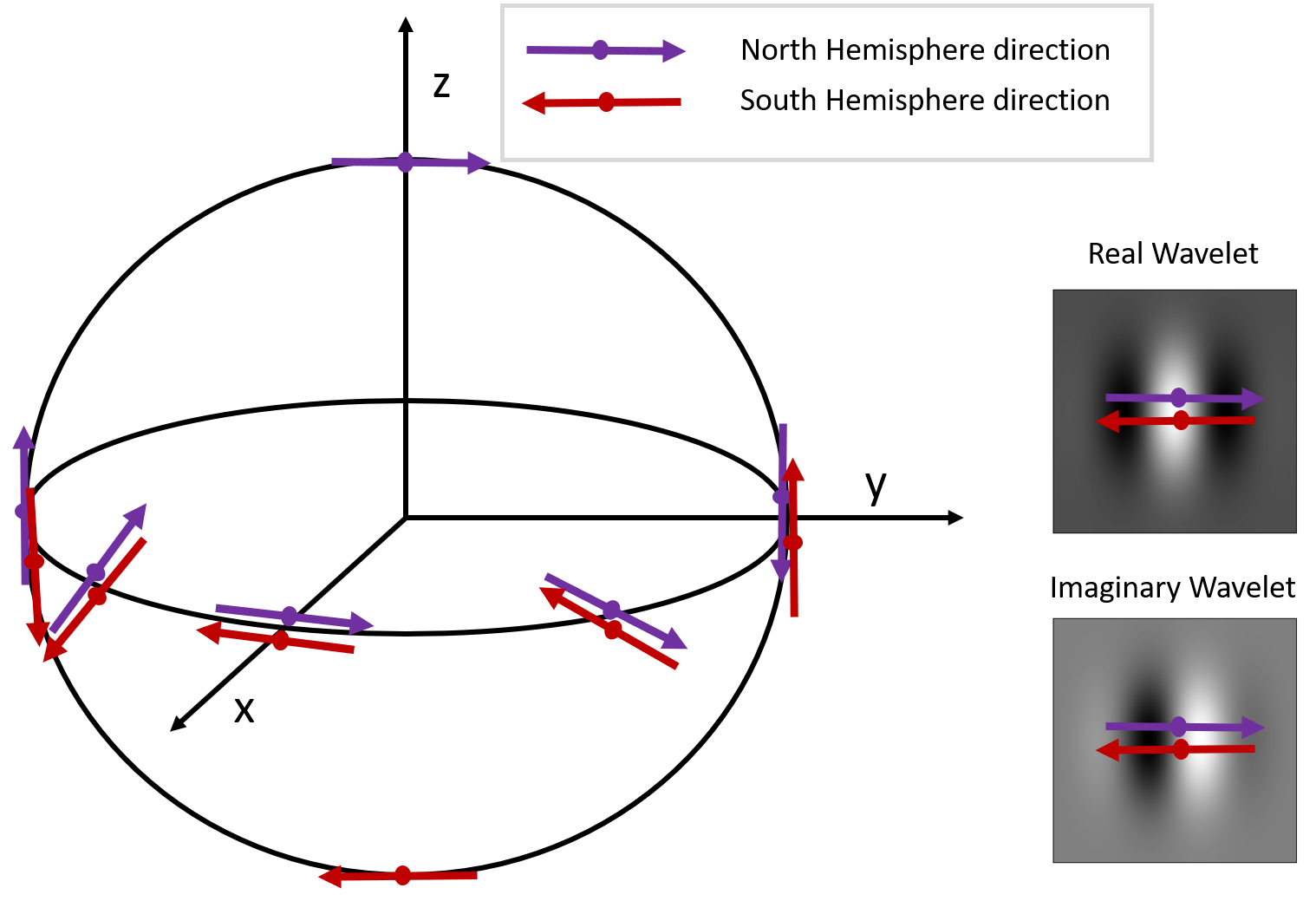}
\caption{Explanation of the local rotational frame utilized to represent the wavelet coefficients. An inversion of direction occurs between the northern (depicted by the blue arrow) and southern (depicted by the red arrow) hemispheres. The effect of this directional inversion is shown in both the real part (top right panel) and the imaginary part (bottom right panel).} \label{fig:wave_dir}
\end{figure}

Although this representation is locally valid, consistent wavelet computation on the sphere requires defining the absolute orientation $\Theta$ and determining meaningful directions. Figure \ref{fig:wave_dir} demonstrates how the wavelet direction is specified. It is important to understand that a consistent orientation definition across the entire sphere is unattainable. The chosen definition perfectly aligns with the symmetric definition for the real part, where orientation does not affect the weights. However, changes in orientation alter the sign of the imaginary part. This definition renders the southern hemisphere definition as the conjugate of the northern hemisphere.

The direction of rotation does not affect the SC statistics, as applying the norm eliminates the influence of this rotation. This is because the SC statistics is computed after normalizing, which conceals the sign of the imaginary component, and also because all calculations are performed at the pixel level, ensuring consistency in the definition.

\subsection{Rotation matrix definition}

The SC coefficients are determined by integrating the local wavelet norm across the entire domain. In instances where there is a preferred orientation in a specific region, typically observed in data retrieved from scanning operations, this preferred orientation influences the data statistics. Consequently, the homogeneity of the entire domain is not ensured, as the dominant orientation at any given scale can vary throughout the domain. To address this, the orientation with the largest amplitude at each pixel is computed at each scale, and this orientation is then referred to as the primary orientation. The SC statistics are then defined relatively to the scanning strategy. To prevent instability, the determination of the maximum is smoothed to avoid rapid changes in the preferred orientation at each pixel. This rotation is applied after calculating the wavelet, resulting in four coefficients per pixel corresponding to the four rotations, which are computed using a precomputed $4\times 4$ matrix for each pixel.

\begin{figure}[]
\centering
\includegraphics[width=0.49\textwidth]{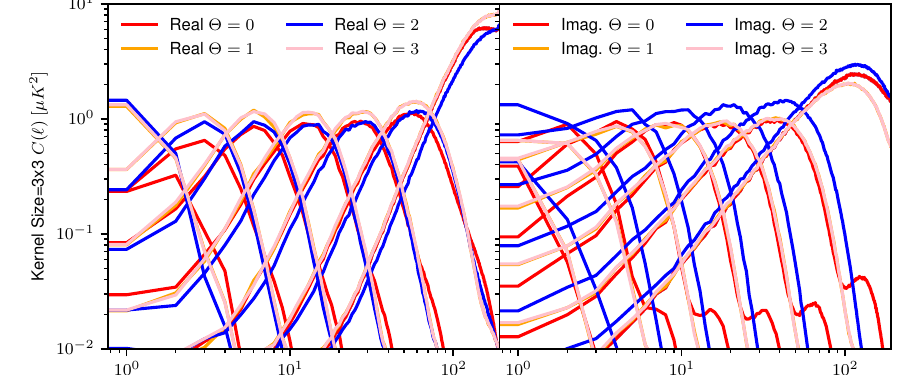}
\includegraphics[width=0.49\textwidth]{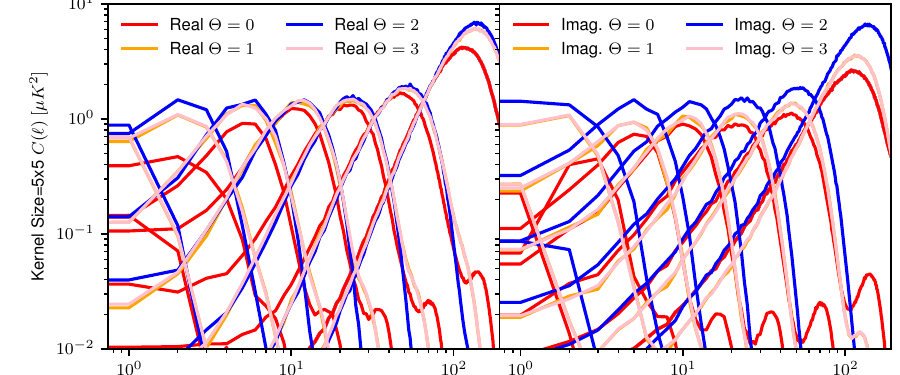}
\caption{Power spectra from the wavelet transformation across the entire sky at nside=64, spanning 7 scales. The upper two panels display results for a wavelet described by a $3\times 3$ kernel size, whereas the lower two panels present identical results for a $5\times 5$ kernel size. The panels on the left depict the power spectra of the real wavelet component, and those on the right illustrate the imaginary component.} \label{fig:WAVE_POWER}
\end{figure}

\begin{figure}[]
\includegraphics[width=0.45\textwidth]{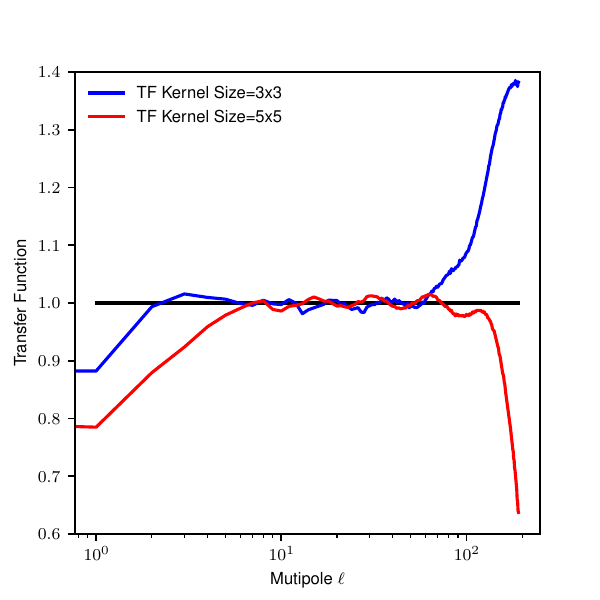}
\caption{Examination of the cumulative transfer function from the wavelet transforms used in this research, derived from 100 samples of white noise. The blue (and red) curve illustrates the aggregate transfer function of the wavelet transforms in all directions and scales, corresponding to a kernel size of $3\times 3$ (and $5\times 5$, respectively).} 
\label{fig:WAVE_TF}
\end{figure}

Figures \ref{fig:WAVE_POWER} present the power spectra of the directive wavelet. Despite the compact size of the kernel used to characterize the wavelet, the resulting power spectra separation is superior to $10^{-2}$. Figure \ref{fig:WAVE_TF} illustrates the complete signal decomposition, displaying the transfer function for the cumulative wavelets. This figure highlights the efficacy of the multiscale decomposition in covering nearly the entire scale domain. A $5\times 5$ kernel faces greater challenges in accurately representing very small multipoles compared to a $3\times 3$ kernel. This issue arises from the longer wavelet length, which requires more pixels for accurate depiction. Specifically, enlarging the mother-wavelet support from $3\times 3$ to $5\times 5$ pixels makes the coarsest dilated kernels span a sizeable fraction of the sky ($\sim 200\deg$).  On the discrete \texttt{HEALPix} grid this wide support is sampled and truncated less accurately (especially near the poles), so the cumulative transfer function drops below unity at \(\ell\lesssim 10\).  On the other hand, a $3 \times 3$ mother-wavelet avoids these issues while still covering the full angular range of interest through dyadic dilations.


\section{Cross scattering covariance statistics}\label{sec:cross_SC}

We report here expressions for the cross SC statistics, extending the ones in Section \ref{sec:scattering_covariance} (Eqs.~\ref{eq:scat_coeff}) to the cross-correlations between two different input fields \( I_1 \) and \( I_2 \):
\begin{equation}\label{eq:cross_scat_coeff}
\begin{aligned}
 & S_1^{\times, \lambda_1} = \Big\langle \Big| \bigl(I_1 \ast \Psi_{\lambda_1} \bigr) \bigl(I_2 \ast \Psi_{\lambda_1} \bigr)^* \Big| \Big\rangle \,,\\
 & S_2^{\times,\lambda_1} = \Big\langle \bigl(I_1 \ast \Psi_{\lambda_1} \bigr) \bigl(I_2 \ast \Psi_{\lambda_1} \bigr)^* \Big\rangle \,, \\
 & S_3^{\times,\lambda_1,\lambda_2} = \mathrm{Cov}\,\Bigl[\,I_1 * \Psi_{\lambda_1},\, \bigl| I_2 * \Psi_{\lambda_2}\bigr| * \Psi_{\lambda_1}\Bigr]\,, \\
 & S_{3p}^{\times,\lambda_1,\lambda_2} = \mathrm{Cov}\,\Bigl[\,I_2 * \Psi_{\lambda_1},\, \bigl| I_1 * \Psi_{\lambda_2}\bigr| * \Psi_{\lambda_1}\Bigr]\,, \\
 & S_4^{\times,\lambda_1,\lambda_2,\lambda_3} = \mathrm{Cov}\,\Bigl[\bigl| I_1 * \Psi_{\lambda_3}\bigr| * \Psi_{\lambda_1},\, \bigl| I_2 * \Psi_{\lambda_2}\bigr| * \Psi_{\lambda_1}\Bigr]\,,
\end{aligned}
\end{equation}
where \(*\) denotes convolution, covariances are defined as $\text{Cov}[X,Y] = \langle XY^* \rangle - \langle X \rangle \langle Y^* \rangle
$ for two complex fields $X$ and $Y$ and \(\langle\,\cdot\rangle\) indicates a spatial average.

\section{SC coefficients, Minkowski functionals and PDFs}\label{sec:extra_validation}

\begin{figure}[!ht]
\centerline{\includegraphics[clip, trim=0.8cm 1.5cm 0.8cm 1.5cm, width=1.07\linewidth]{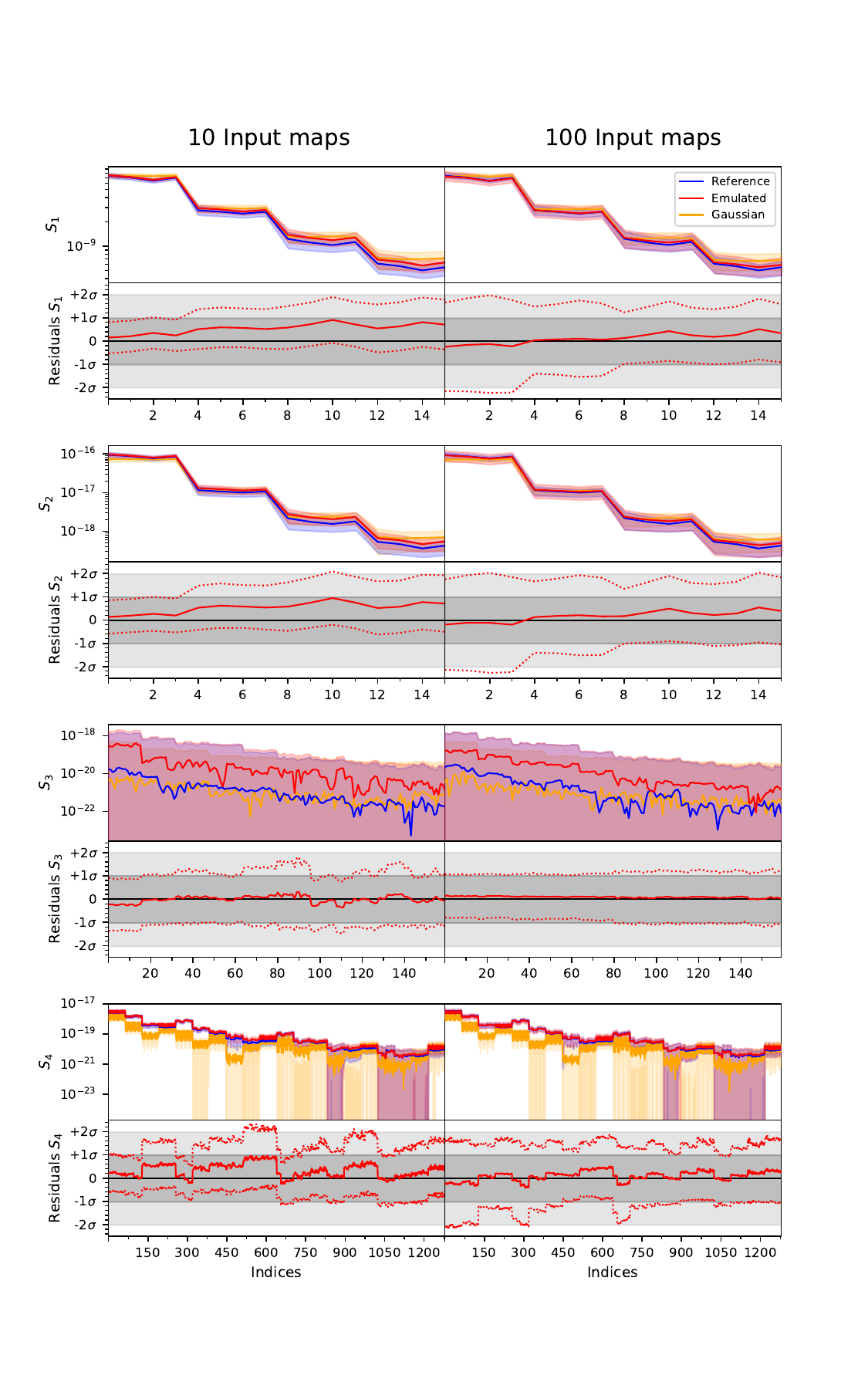}}
\caption{SC coefficients of the $Q$ map for the reference (blue), emulated (red) and Gaussian (orange) datasets, for both the 10-maps (left panels) and 100-maps (right panels) input datasets. In each row, the top panel displays the average SC coefficient with 1-$\sigma$ shading, while the bottom panel shows the residual, $\Delta S_{X} = \left(\langle S_{X}^\mathrm{emu}\rangle - \langle S_{X}^\mathrm{ref}\rangle \right)/\sigma_{S_{X}}^\mathrm{ref}$ (solid red) along with $\Delta S_{X}^\pm = \left(\langle S_{X}^\mathrm{emu}\rangle - \langle S_{X}^\mathrm{ref}\rangle \pm \sigma_{S_{X}}^\mathrm{emu}\right)/\sigma_{S_{X}}^\mathrm{ref}$ (dotted red), where $\sigma_{S_{X}}^\mathrm{ref}$ is the standard deviation from the reference set and $\sigma_{S_{X}}^\mathrm{emu}$ is the emulated dataset's one.}
\label{fig:scat_coeff}
\end{figure}

\begin{figure}[!ht]
\centerline{\includegraphics[width=1.\linewidth]{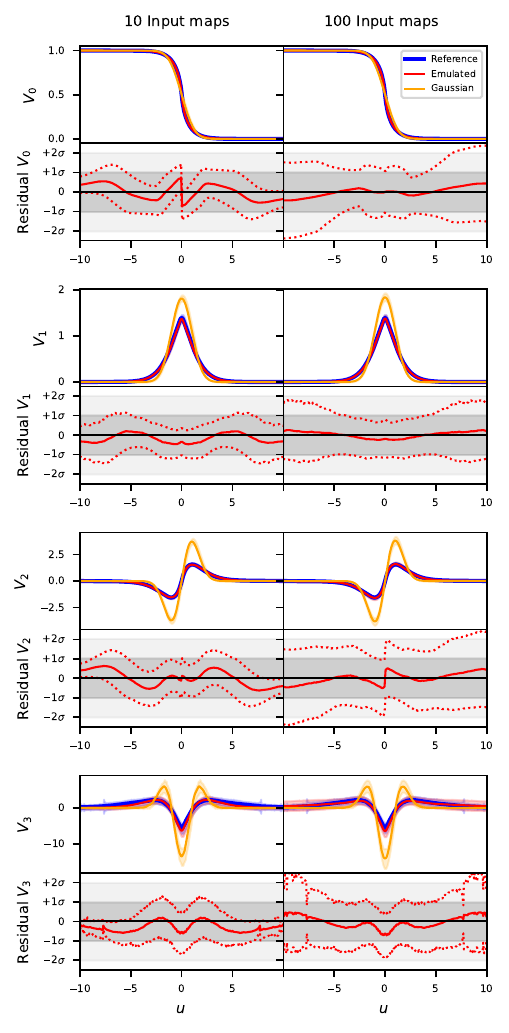}}
\caption{The four SO(3) Minkowski functionals for the reference (blue), emulated (red) and Gaussian (orange) datasets, for both the 10-maps (left panels) and 100-maps (right panels) input datasets. In each row, the top panel displays the average Minkowski functional with 1-$\sigma$ shading, while the bottom panel shows the residual, $\Delta V_{X} = \left(\langle V_{X}^\mathrm{emu}\rangle - \langle V_{X}^\mathrm{ref}\rangle \right)/\sigma_{V_{X}}^\mathrm{ref}$ (solid red) along with $\Delta V_{X}^\pm = \left(\langle V_{X}^\mathrm{emu}\rangle - \langle V_{X}^\mathrm{ref}\rangle \pm \sigma_{V_{X}}^\mathrm{emu}\right)/\sigma_{V_{X}}^\mathrm{ref}$ (dotted red), where $\sigma_{V_{X}}^\mathrm{ref}$ is the standard deviation from the reference set and $\sigma_{V_{X}}^\mathrm{emu}$ is the emulated dataset's one. }
\label{fig:minkowski}
\end{figure}

\begin{figure}[!ht]
\centerline{\includegraphics[width=1.\linewidth]{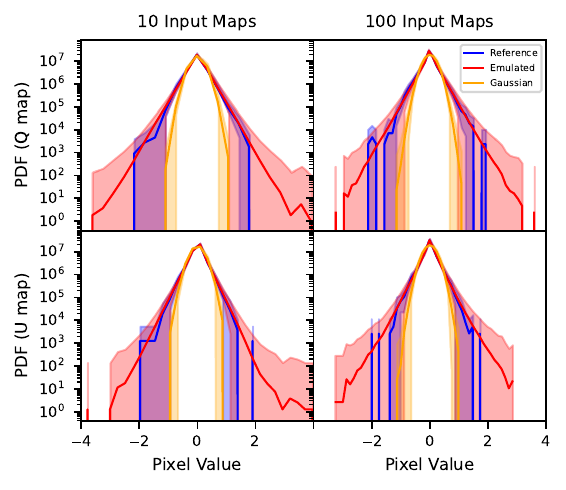}}
\caption{PDFs in pixel space for the maps in the reference (blue), emulated (red) and Gaussian (orange) datasets, for both the 10-maps (left panels) and 100-maps (right panels) input datasets. Top panels show the $Q$ maps, while bottom ones the $U$ maps. The shaded area represents the standard deviation of the maps PDFs.}
\label{fig:pdfs}
\end{figure}

We complement the validation shown in Section \ref{sec:results} by plotting the SC coefficients (Fig.~\ref{fig:scat_coeff}), the Minkowski functionals (Fig.~\ref{fig:minkowski}) and the PDFs (probability density functions) in pixel space (Fig.~\ref{fig:pdfs}) of all the datasets considered here. As expected, we find remarkable agreement in all these test statistics.

Figure~\ref{fig:scat_coeff} shows the SC coefficients for the reference (blue), emulated (red) and Gaussian (orange) datasets, for both the 10-maps (left panels) and 100-maps (right panels) input datasets. Each panel shows the average SC coefficient surrounded by a shaded area representing the standard deviation of the coefficients. The indices are ordered from the smallest to the largest convolved scales for each coefficient.
To quantify differences between the SC coefficients of the reference and emulated datasets, we define their residuals as 
\[
\Delta S_{X} \;=\; \frac{\langle S_{X}^\mathrm{emu}\rangle \;-\; \langle S_{X}^\mathrm{ref}\rangle}{\sigma_{S_{X}}^\mathrm{ref}},
\]  
where $X=1,2,3,4$ identifies the specific SC coefficient, \(\langle S_{X}^\mathrm{emu}\rangle\) and \(\langle S_{X}^\mathrm{ref}\rangle\) are the average coefficients in the emulated and reference sets, respectively, and \(\sigma_{S_{X}}^\mathrm{ref}\) is the standard deviation measured in the reference set. As shown in the bottom panels of Fig.~\ref{fig:scat_coeff} (solid red lines), \(\Delta S_{X}\) remains well within the 1\(\sigma\) band for all indices. 
To compare the variance in the two datasets, we also plot  
\[
\Delta S_{X}^\pm \;=\; \frac{\langle S_{X}^\mathrm{emu}\rangle \;-\; \langle S_{X}^\mathrm{ref}\rangle \;\pm\; \sigma_{S_{X}}^\mathrm{emu}}{\sigma_{S_{X}}^\mathrm{ref}},
\]  
(dotted red lines), where \(\sigma_{S_{X}}^\mathrm{emu}\) is the standard deviation in the emulated dataset. 
We note that the average emulated $S_4$ coefficients closely follow the reference set, suggesting that this coefficient is among the most dominant ones in driving the gradient descent. Moreover, $S_4$ is the coefficient showing the most deviation from the Gaussian dataset, suggesting that $S_4$ is also the main driver behind the success of the generative model over the naive Gaussian augmentation from the angular power spectrum of the input maps. On the other hand, $S_3$ appears to contain limited information about the field under study: it is very small in amplitude and weakly constrained in this specific case, oscillating widely but still within the 1-$\sigma$ band.

In the 10-input maps scenario (left panels), the SC coefficient variance at the smallest-convolved scales is slightly underestimated for all but \(S_3\), whereas in the 100-input maps case it is overestimated (again with the exception of \(S_3\)). We attribute this behavior to the different targets used during gradient descent, namely the average-target in the 10-maps setup versus the single-target in the 100-maps setup. Specifically, employing a single-target strategy overestimates the SC coefficient variance by incorporating both the “within-batch” variance introduced by mean-field gradient descent for each sampled target, and the “within-target” variance explained by the different targets in the input dataset. Despite this additional variance in the SC coefficients, the other key metrics presented in Section \ref{sec:results} are unaffected by the overestimation. Nonetheless, we expect that this excess variance could be mitigated by adding a suitable term in the loss function to penalize large “within-batch” variance. We leave further exploration of this approach to future work.

We compute the four Minkowski functionals $V_1$, $V_2$, $V_3$ and $V_4$ of the polarization maps (Fig.~\ref{fig:minkowski}) in the SO(3) formalism, which is the most appropriate for $Q$ and $U$ polarization maps \citep{CarronDuque:2023bph}. For this purpose, We use the \texttt{pynkowski} software\footnote{\href{https://github.com/javicarron/pynkowski}{https://github.com/javicarron/pynkowski}} \citep{Carones:2022rbv, CarronDuque:2023bph}. Minkowski functionals (hereafter MFs) provide a compact and computatonally inexpensive summary of geometrical and topological information of the field (such as area, boundary lenght and Euler characteristic or genus as a function of a threshold) and are sensitive to long-range properties of the field and higher-order correlations not directly constrained by the power spectrum and SC coefficients. MFs represent therefore a complementary summary statistics, and demonstrating that the emulator reproduces the MFs of the reference set is thus a stringent test for our intended applications. Similarly to the SC coefficients, we quantify differences between the Minkowski functionals of the reference and emulated datasets, by defining their residuals as 
\[
\Delta V_{X} \;=\; \frac{\langle V_{X}^\mathrm{emu}\rangle \;-\; \langle V_{X}^\mathrm{ref}\rangle}{\sigma_{V_{X}}^\mathrm{ref}},
\]  
where $X=1,2,3,4$ identifies the specific Minkowski functional, \(\langle V_{X}^\mathrm{emu}\rangle\) and \(\langle V_{X}^\mathrm{ref}\rangle\) are the average functionals in the emulated and reference sets, respectively, and \(\sigma_{V_{X}}^\mathrm{ref}\) is the standard deviation measured in the reference set.  
We also compare the variance in the two datasets by plotting  
\[
\Delta V_{X}^\pm \;=\; \frac{\langle V_{X}^\mathrm{emu}\rangle \;-\; \langle V_{X}^\mathrm{ref}\rangle \;\pm\; \sigma_{V_{X}}^\mathrm{emu}}{\sigma_{V_{X}}^\mathrm{ref}},
\]  
(dotted red lines), where \(\sigma_{V_{X}}^\mathrm{emu}\) is the standard deviation in the emulated dataset. 
The bottom panels of Fig.~\ref{fig:minkowski} (solid red lines), show that \(\Delta V_{X}\) remains well within the 1\(\sigma\) band across the threshold values $u$, although some residuals features are indeed prominent. 
Specifically, the feature around $u=0$ (especially pronounced in $V_0$ in the $N_{\rm input}=10$ case and $V_2$ for $N_{\rm input}=100$) is a known effect of thresholding the MFs at such low-values, making the variance extremely small in the neighbourhood of $u=0$ and therefore creating a spike in the ratio \(\Delta V_{X}\). Moreover, the MFs are higher-order, non-linear statistics and therefore converge more slowly for any finite ensemble: this might be the cause of the pronounced wiggles in the residuals.

We also show the maps PDFs in pixel space for the reference (blue), emulated (red) and Gaussian (orange) datasets, for both the 10-maps (left panels) and 100-maps (right panels) input datasets (Fig.~\ref{fig:pdfs}). The top panels show the $Q$ maps, while bottom ones the $U$ maps. Again, the shaded area represents the standard deviation of the maps PDFs. Also here the emulated and reference maps appear to be remarkably consistent.

\section{A note on the reference and input datasets}\label{sec:appendix_Gaussian}

In this appendix, we offer additional clarification and motivation for the decision to construct the reference and the two input datasets by drawing Gaussian random realizations from the pixel-to-pixel covariance matrix of the original dataset produced from the TOD simulations (i.e. the reference covariance matrix described in Section \ref{sec:simulations}), rather than using the original maps directly.

First, the empirical covariance matrix built from a finite dataset of simulations is plagued by sample variance. While this estimator is unbiased, the inverse of the empirical covariance -- used for example in the $\chi^2$ computation (such as the one in Eq.~\ref{eq:chi-square}) and in likelihoods -- is biased \citep{Hartlap:2006kj}. Although in principle techniques such as the Hartlap factor \citep{Hartlap:2006kj} or the Ledoit–Wolf correction \citep{ledoit} can mitigate this, their performance can be difficult to evaluate, and they require additional modeling assumptions. Because one of our key metrics for assessing the performance of the emulator is its ability to reproduce the covariance matrix itself (alongside the \(\chi^2\) in pixel space), it is crucial for us to have access to a ground-truth reference covariance matrix. Therefore, instead of emulating directly the original maps, we generate a new reference dataset—together with two input datasets of 10 and 100 maps, respectively—using the covariance matrix of the original dataset. This ensures that the reference covariance matrix serves as the ground truth for all these datasets.

Second, the probability density function of each pixel in the original maps is already close to Gaussian, due to the way these maps were generated (i.e adding random Gaussian fluctuations at the TOD level; see Section \ref{sec:simulations}). In our specific case, the pixel-to-pixel spatial information in the original dataset appears to be very well encoded in the reference covariance matrix: Gaussian realizations drawn from this covariance yield both maps and summary statistics (such as power spectrum, PDFs, Minkowski functionals and SC coefficients) very compatible with those of the original dataset. Therefore, no substantial difference in our results is expected when using the original dataset in lieu of the Gaussian realizations.

Third, generating Gaussian realizations from the reference covariance matrix enables us to create datasets that are completely independent, in a statistical sense, from the simulations used to estimate that same reference covariance. In high-accuracy analyses, even minimal overlap between the simulations used to build the covariance matrix and those considered as ``data'' (see for example Eq.~\ref{eq:chi-square}) can, in principle, artificially suppress sampling variance and, consequently, the bias in the inverse covariance estimator. Our approach avoids this potential problem without having to split the original dataset into a set used for covariance matrix estimation and a set used only for validation, allowing us to use all 10,000 simulations in the original dataset to estimate the reference covariance. 

\end{appendix}

\end{document}